\documentclass[fleqn,usenatbib]{mnras}

\usepackage{newtxtext,newtxmath}

\usepackage[T1]{fontenc}

\DeclareRobustCommand{\VAN}[3]{#2}
\let\VANthebibliography\thebibliography
\def\thebibliography{\DeclareRobustCommand{\VAN}[3]{##3}\VANthebibliography}

\usepackage{graphicx}	
\usepackage{amsmath}	
\usepackage{subfigure}
\usepackage{natbib}
\usepackage{color}
\usepackage{tabularx}
\usepackage{longtable}
\usepackage{bm}
\usepackage{threeparttable}
\usepackage[normalem]{ulem}
\usepackage{verbatim}

\newcommand{\todo}[1]{\textcolor{red}{#1}}

\newcommand{\code}[1]{{\texttt{#1}}}
\newcommand{\tess}{{\it TESS}}
\newcommand{\kepler}{{\it Kepler}}

\newcommand{\gaia}{{\it Gaia}}
\newcommand{\wise}{{\it WISE}}

\newcommand{\tar}{{TOI-530}}

\title[TOI-530b]{TOI-530b: A giant planet transiting an M dwarf detected by TESS}

\author[T. Gan et al.]{Tianjun Gan,$^{1}$\thanks{E-mail: gtj18@mails.tsinghua.edu.cn}
Zitao Lin,$^{2}$
Sharon~Xuesong~Wang, $^{1}$\thanks{E-mail: sharonw@mail.tsinghua.edu.cn}
Shude Mao, $^{1,3}$
Pascal Fouqu\'e,$^{4,5}$
Keivan G.\ Stassun, $^{6,7}$
\newauthor
Steven Giacalone, $^{8}$
Akihiko Fukui, $^{9,10}$
Felipe Murgas, $^{10,11}$
David R. Ciardi, $^{12}$
Steve~B.~Howell, $^{13}$
\newauthor
Karen A.\ Collins, $^{14}$
Avi Shporer, $^{15}$
Luc Arnold, $^{4}$
Thomas~Barclay, $^{16,17}$
David Charbonneau, $^{14}$
\newauthor
Jessie Christiansen, $^{12}$
Ian~J.~M.~Crossfield, $^{18}$
Courtney~D.~Dressing, $^{8}$
Ashley~Elliott, $^{19}$
\newauthor
Emma Esparza-Borges, $^{10,11}$
Phil Evans,$^{20}$
Crystal~L.~Gnilka, $^{13}$
Erica~J.~Gonzales, $^{21}$
Andrew W. Howard, $^{22}$
\newauthor
Keisuke Isogai, $^{23,24}$
Kiyoe Kawauchi, $^{24}$
Seiya Kurita, $^{25}$
Beibei Liu, $^{26}$
John H. Livingston, $^{27}$
\newauthor
Rachel A.~Matson, $^{28}$
Norio Narita, $^{9,10,29,30}$
Enric Palle, $^{10,11}$
Hannu Parviainen, $^{10,11}$
\newauthor
Benjamin~V.~Rackham, $^{15,31,\ast}$
David R. Rodriguez, $^{32}$
Mark Rose, $^{13}$
Alexander Rudat, $^{15}$
Joshua~E.~Schlieder, $^{16}$
\newauthor
Nicholas~J.~Scott, $^{13}$
Michael Vezie, $^{15}$
George~R.~Ricker,$^{15}$
Roland~Vanderspek,$^{15}$
David~W.~Latham,$^{14}$
\newauthor
Sara~Seager,$^{15,31,33}$
Joshua~N.~Winn,$^{34}$
and Jon~M.~Jenkins$^{13}$
\\
Affiliations are listed at the end of the paper
}

\date{Accepted XXX. Received YYY; in original form ZZZ}

\pubyear{2021}

\begin{document}
\label{firstpage}
\pagerange{\pageref{firstpage}--\pageref{lastpage}}
\maketitle

\begin{abstract}
We report the discovery of \tar b, a transiting giant planet around an M0.5V dwarf, delivered by the Transiting Exoplanet Survey Satellite (\tess). The host star is located at a distance of $147.7\pm0.6$ pc with a radius of $R_{\ast}=0.54\pm0.03\ R_{\odot}$ and a mass of $M_{\ast}=0.53\pm0.02\ M_{\odot}$. We verify the planetary nature of the transit signals by combining ground-based multi-wavelength photometry, high resolution spectroscopy from SPIRou as well as high-angular-resolution imaging. With $V=15.4$~mag, \tar b is orbiting one of the faintest stars accessible by ground-based spectroscopy. Our model reveals that \tar b has a radius of $0.83\pm0.05\ R_{J}$ and a mass of $0.4\pm0.1\ M_{J}$ on a $6.39$-d orbit. \tar b is the sixth transiting giant planet hosted by an M-type star, which is predicted to be infrequent according to core accretion theory, making it a valuable object to further study the formation and migration history of similar planets. We discuss the potential formation channel of such systems. 
\end{abstract}




\section{Introduction}

M dwarfs are popular targets for exoplanet research. First, radial velocity (RV) variations induced by the planets around M dwarfs are more significant than those around solar-like stars, making it possible to obtain precise mass measurement towards the terrestrial planet end of the mass distribution. Second, their small stellar radii lead to a large planet-to-star radius ratio, which favors transit detections and further photometric follow-up observations. Planets around M dwarfs are also attractive sources for atmospheric characterization through transmission or emission spectroscopy \citep{Kempton2018,Batalha2018} as they yield a higher signal-to-noise ratio than equivalent systems with other types of hosts (e.g., LHS 3844b, \citealt{Vanderspek2019,Kreidberg2019,Diamond2020}). Finally, due to the low stellar luminosity (typically $L<0.1\ L_{\odot}$), the habitable zone of M dwarfs is closer to the host star when compared with luminous stars (e.g., TOI-700d, \citealt{Gilbert2020,Rodriguez2020}), which offers particular advantages to look for planets with potential biosignatures. 

Over the last two decades, more than a thousand transiting giant planets (defined as $M_{p}>0.3\ M_{J}$) have been discovered thanks to successful ground-based surveys, including HATNet \citep{Bakos2004}, SuperWASP \citep{Pollacco2006}, KELT \citep{Pepper2007,Pepper2012} and NGTS \citep{Chazelas2012,Wheatley2018} as well as space transit missions like {\it CoRoT} \citep{Baglin2006}, {\it Kepler} \citep{Borucki2010} and {\it K2} \citep{Howell2014}. However, even though M dwarfs are the most abundant stellar population in our Milky Way \citep{Henry2006}, only five giant planets have been confirmed to transit them: Kepler-45b \citep{Johnson2012}, HATS-6b \citep{Hartman2015}, NGTS-1b \citep{Bayliss2018}, HATS-71b \citep{Bakos2020} and TOI-1899b \citep{Canas2020}. The deficiency of such systems is thought to be caused by the failed growth of a massive core to start runaway accretion before the gaseous protoplanetary disk dissipates due to the low surface density \citep{Laughlin2004,Ida2005,Kennedy2008,Liu2020}. Indeed, previous statistical studies of the occurrence rates from \kepler\ show that planets with radii between $1R_{\oplus}$ and $4R_{\oplus}$ are frequent around low-mass stars \citep{Dressing2013,Dressing2015,Hardegree-Ullman2019}. Some of these small planets are possibly the bare cores of failed gas giants. Nevertheless, microlensing surveys have found plenty of cold Jupiters ($a\gtrsim 1$ AU) around M dwarfs (e.g., \citealt{Zang2018}), which hints that outer giant planets are probably not rare. A handful of such cases have also been reported by long-term RV observations (e.g., GJ 876b, \citealt{Marcy2001}; GJ 849b, \citealt{Butler2006}; GJ 179b, \citealt{Howard2010}; HIP 79431b, \citealt{Apps2010}). Gravitational instability is speculated to be the alternative formation mechanism responsible for the surprising number of long-period gas giants around M dwarfs \citep{Boss2000,Morales2019}. But it is still unclear how these short-period ($P\lesssim 30$ d) transiting gas giants were formed, and whether they have migrated into their current orbits due to the lack of such systems. Therefore, establishing a well-characterized sample of this kind of planet is an important step to study their formation. Further Rossiter-McLaughlin \citep{Rossiter1924,McLaughlin1924} or Doppler tomography \citep{Marsh2001} measurements could reveal the obliquity of these systems, providing important clues about the dynamical history of the planets (e.g., \citealt{Albrecht2012}).

The Transiting Exoplanet Survey Satellite (\tess, \citealt{Ricker2014,Ricker2015}), which has performed a two-year all-sky survey, offers exciting opportunities to increase the number of transiting giant planets around M dwarfs. Although \tess\ has already identified several such planet candidates, the intrinsic faintness of their hosts ($V\gtrsim15$ mag) challenges most ground-based optical spectroscopic facilities to further conduct detailed RV follow-up observations. Some efforts have already been made to validate those planets through multi-color transit modeling and phase curve analysis (e.g., TOI-519b, \citealt{Parviainen2021}). The new-generation near-infrared spectrograph SPIRou on the Canada-France-Hawaii-Telescope (CFHT) opens a new window to characterize planets around faint stars \citep{Artigau2014,Donati2020}. It was designed to perform high-precision velocimetry and spectropolarimetry studies. Early observations from SPIRou have shown that it can reach $2\sim10$ m/s precision for stars with $H<10$ mag \citep{Moutou2020,Klein2021,Artigau2021}. Although simulations predict that SPIRou could reach $<2$ m/s RV precision for inactive M dwarfs with $J<10$ mag (see Figure 5 in \citealt{Cloutier2018}), precision for faint stars has yet to be determined observationally.

Here we report the discovery of a new transiting giant planet around an M-dwarf star \tar. We present RV measurements from SPIRou that allow us to obtain a precise companion mass and thus confirm its planetary nature. The rest of the paper is organized as follows. We describe all space and ground-based observational data used in this work in Section \ref{observations}. Section \ref{stellar_properties} presents the stellar properties. In Section \ref{analysis}, we show our analysis of the light curves and RV data. We discuss the prospects of future atmospheric characterization of \tar b and its potential formation channel in Section \ref{discussion}. We conclude with our findings in Section \ref{conclusion}.






\section{Observations}\label{observations}
\subsection{\tess\ photometry}
\begin{figure*}
\centering
\includegraphics[width=0.94\columnwidth]{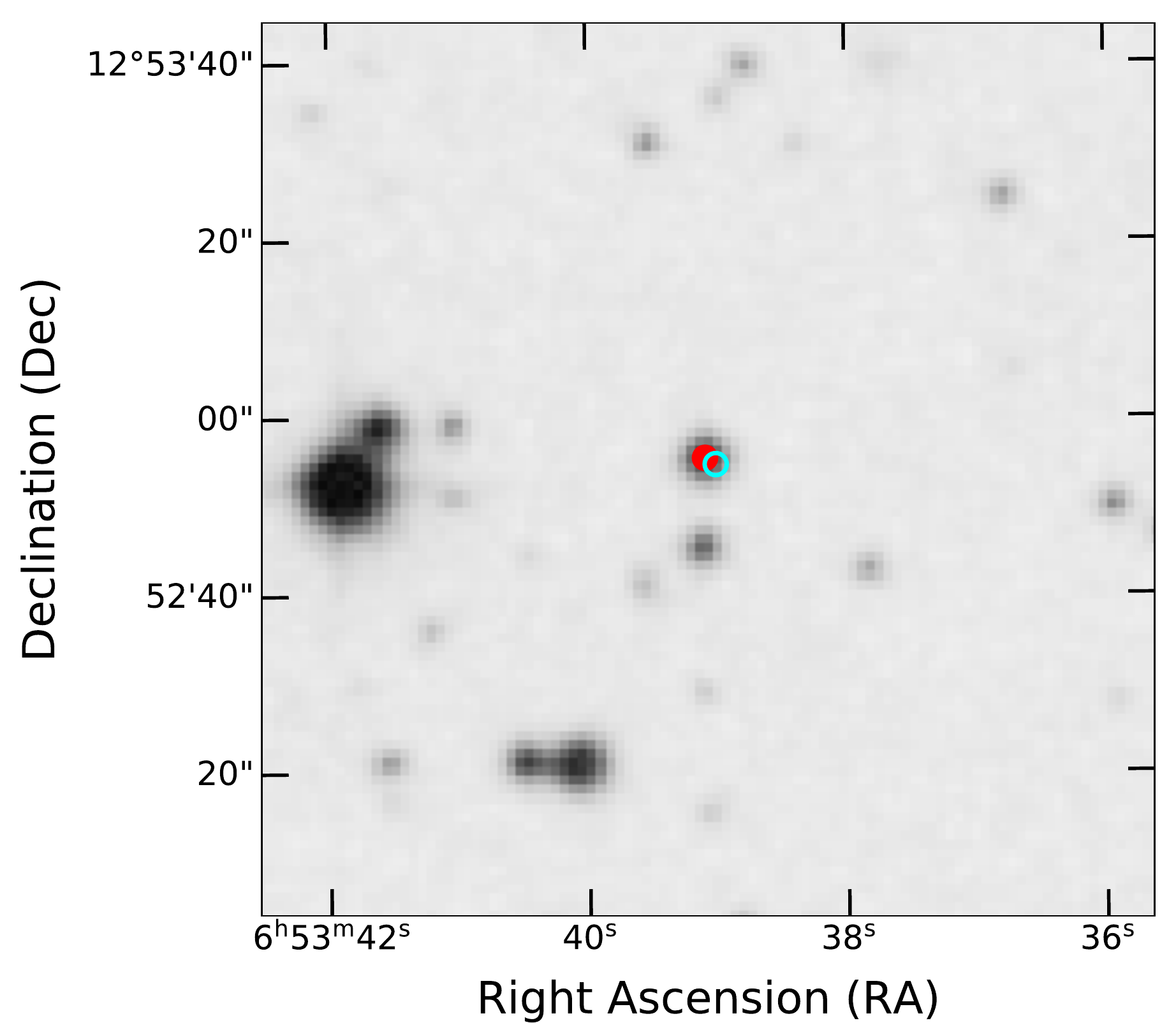}
\includegraphics[width=1.06\columnwidth]{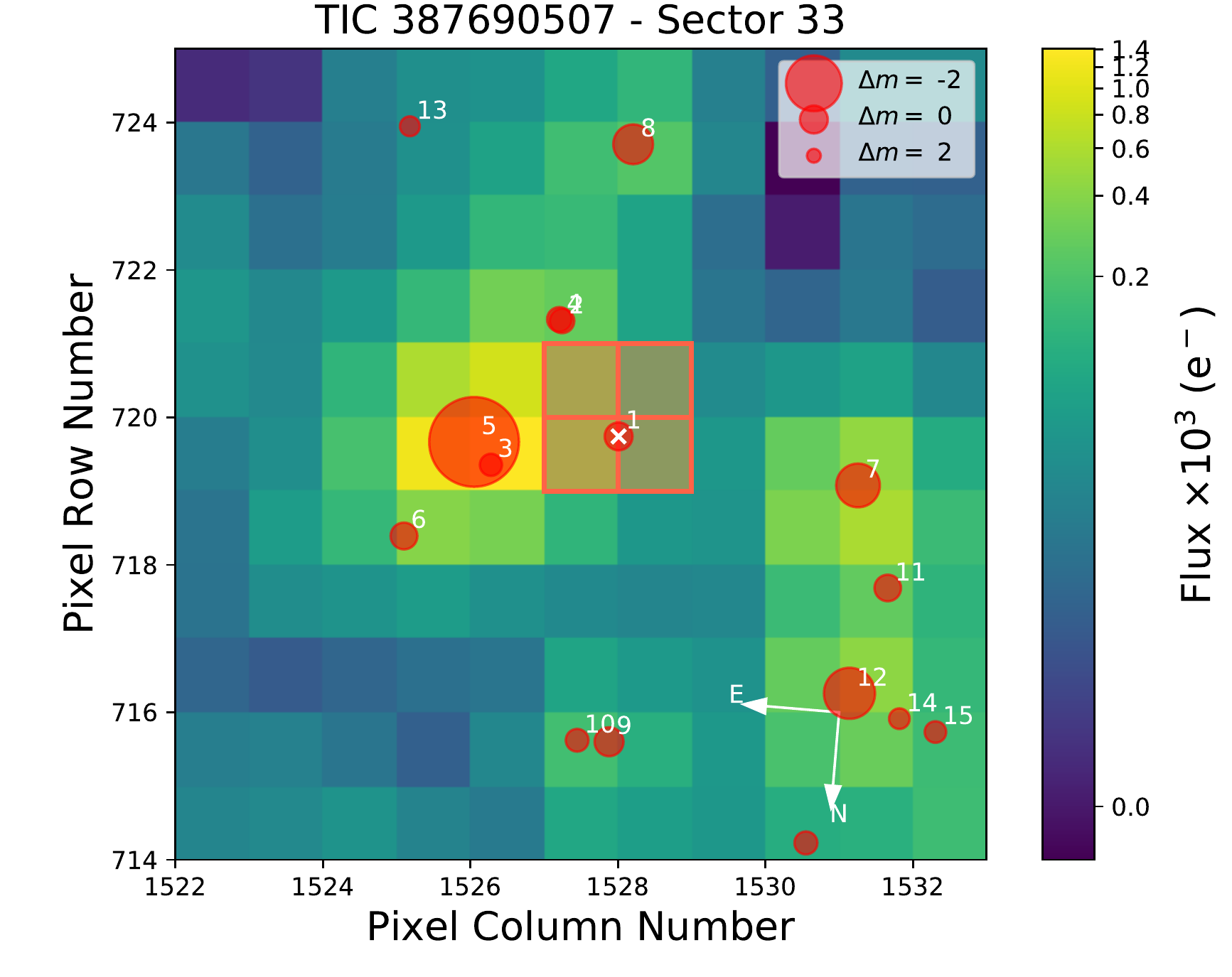}
\caption{{\it Left panel}: The POSS2 blue image of \tar\ taken in 1996. The center red dot is the target star in this image and the cyan circle shows its current position, which rules out the unassociated distant eclipsing binary scenario. {\it Right panel}: Target pixel file (TPF) of \tar\ in \tess\ Sector 33 (created with \code{tpfplotter}, \citealt{Aller2020}). Different sizes of red circles represent different magnitudes in contrast with \tar\ ($\Delta m$). The aperture used to extract the photometry is overplotted with a red-square region.} 
\label{fov}
\end{figure*}

\tar\ was observed by \tess\ on its Camera 1 with the two-minute cadence mode in Sector 6 during the primary mission and Sector 33 during the extended mission. The current data span from 2018 December 15th to 2021 January 13th, consisting of 14830 and 17547 measurements, respectively. The target will be revisited in Sectors 44-45 between 2021 October 12th and 2021 December 2nd. Figure \ref{fov} shows the POSS2 and \tess\ images centered on \tar.

The photometric data from Sector 6 were initially reduced by the Science Processing Operations Center (SPOC; \citealt{Jenkins2016}) pipeline, developed based on the \kepler\ mission's science pipeline. The simple aperture photometry (SAP) flux time series was corrected for instrumental and systematic effects, and for crowding and dilution with the Presearch Data Conditioning (PDC; \citealt{Stumpe2012,Smith2012,Stumpe2014}) module. Transit signals were searched using the Transiting Planet Search \citep[TPS;][]{Jenkins2002,Jenkins2017} algorithm on 17 February 2019, yielding a strong transit signal at a period of $\sim$6.39 days and a transit duration of $\sim$2.5 hours. The transit signature and pixel data passed all the  validation tests \citep{Twicken2018,Li2019,Guerrero2021}, including locating the source of the transit signature to within 1 - 3 arcsec of the target star, and no further transiting planet signatures were identified in a search of the residual light curve. The vetting results were reviewed by the TESS Science Office (TSO) and issued an alert for \tar b as a planet candidate on 28 March 2019.

\begin{figure*}
\centering
\includegraphics[width=\textwidth]{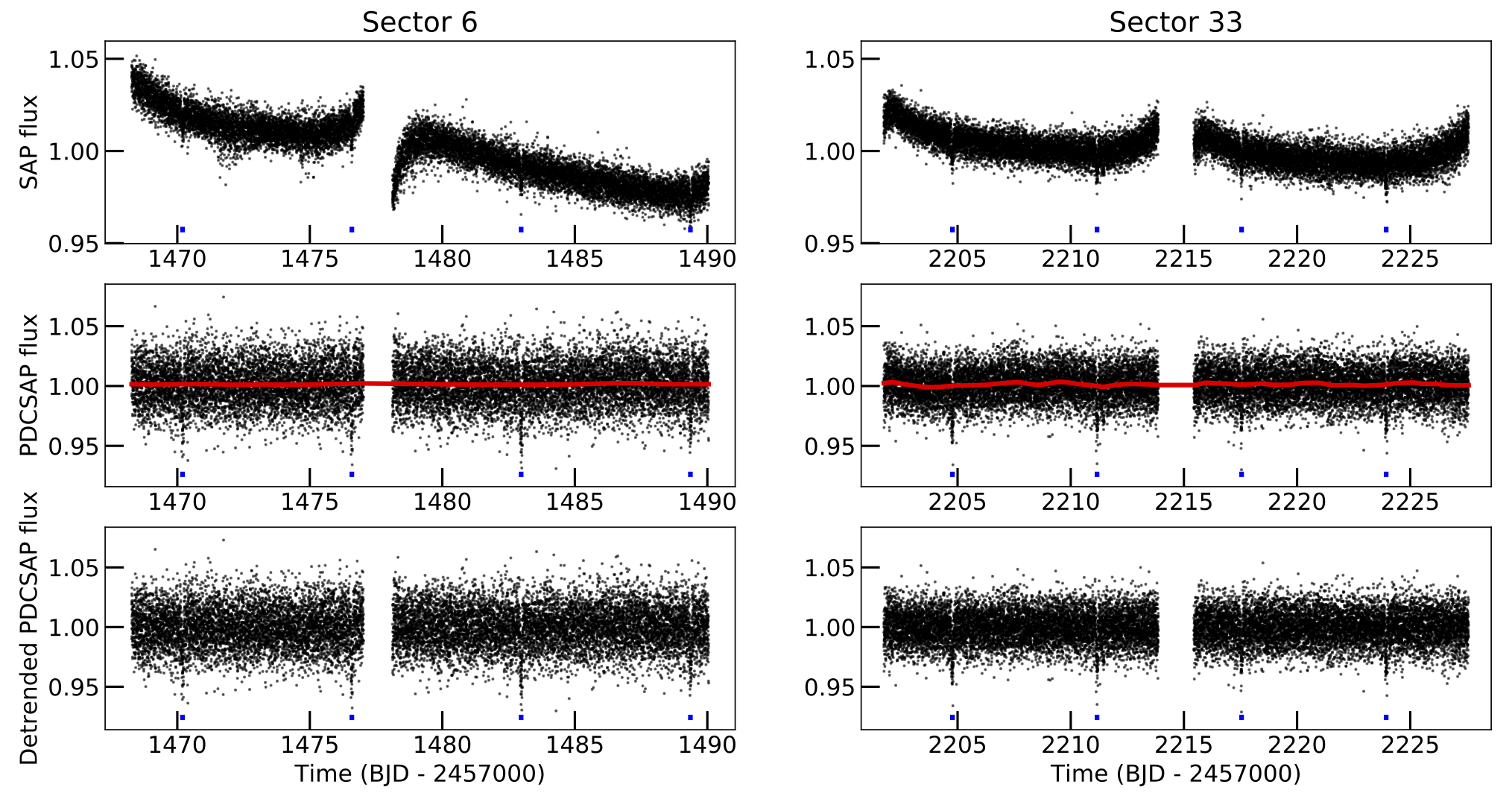}
\caption{{\it Top panels}: The original \tess\ SAP light curves of \tar\ from Sector 6 and 33. {\it Middle panels}: The PDCSAP light curves of \tar\ along with the best-fit GP model shown as red solid lines. {\it Bottom panels}: The detrended PDCSAP light curves. The transits of \tar b are marked in blue ticks.} 
\label{transit_detrend}
\end{figure*}

We downloaded the Presearch Data Conditioning Simple Aperture Photometry (PDCSAP) light curve from the Mikulski Archive for Space Telescopes\ (MAST\footnote{\url{http://archive.stsci.edu/tess/}}) using the \code{lightkurve} package \citep{lightkurvecollaboration,lightkurve}. Combining the datasets of two sectors, we conducted an independent transit search by utilizing the Transit Least Squares (TLS; \citealt{Hippke2019}) algorithm, which is an advanced version of Box Least Square (BLS; \citealt{Kovacs2002}), after smoothing the full light curve with a median filter. We recovered the 6.387 d transits with a signal detection efficiency (SDE) of $\sim50$. After subtracting the TLS model from the \tess\ data, we did not find any other significant transit signals existing in the light curve. We detrended the raw \tess\ light curve by fitting a Gaussian Process (GP) model with a Mat\'{e}rn-3/2 kernel using the \code{celerite} package \citep{Foreman2017}, after masking out all in-transit data. We show the reprocessed light curve in Figure \ref{transit_detrend}. 


\subsection{Ground-Based photometry}\label{gbp}
We collected a series of ground-based observations of \tar, as part of the TESS Follow-up Observing Program (TFOP\footnote{\url{https://tess.mit.edu/followup}}), to (1) confirm the transit signal on target and rule out nearby eclipsing binary scenario; (2) examine the chromaticity; and (3) refine the transit ephemeris and radius measurement. These observations were scheduled with the help of the \tess\ Transit Finder (\code{TTF}), which is a customized version of the \code{Tapir} software package \citep{Jensen2013}. Due to the observational constraints, unfortunately, we only covered the egress of the event. We summarize the details in Table \ref{po} and describe individual observations below. We show the raw and detrended ground-based light curves in Figure \ref{ground_transit_detrend} (see Section \ref{joint_fit_TESS_ground}).

\subsubsection{El Sauce}\label{el}
An egress was observed on UT 2019 November 21 in the $R_{c}$ band using the Evans telescope (0.36 m) at the El Sauce Observatory, Chile. The STT 1603 camera has a pixel scale of $\rm 1.47''$ per pixel. We acquired a total of 65 images over 205 minutes. Photometric analysis was carried out using \code{AstroImageJ} \citep{Collins2017} with an uncontaminated aperture of $5.88''$. We excluded all nearby stars within $1'$ as the source causing the \tess\ signal with brightness difference down to $ \Delta T \sim 4.1$ mag, and confirmed the signal on target.

\subsubsection{MuSCAT2}\label{muscat2}
We observed an egress of TOI-530b on the night of UT 2020 January 4 with the multicolor imager MuSCAT2 \citep{2019JATIS...5a5001N} mounted on the 1.52 m Telescopio Carlos S\'{a}nchez at Teide Observatory, Tenerife, Spain. MuSCAT2 has a field of view of $7.4' \times 7.4'$ with a pixel scale of $0.44''$ pixel$^{-1}$ and is able to obtain simultaneous photometry in four bands ($g$, $r$, $i$, and $z_s$). The observations were made with the telescope in optimal focus and the exposure times for each band were 45 s for $g$, 30 s for $r$ and $i$, and 20 s for $z_s$ band. The data were calibrated using standard procedures (dark and flat calibration). Aperture photometry and transit light curve fit was performed using MuSCAT2 pipeline (\citealp{2020A&A...633A..28P}); the pipeline finds the aperture that minimizes the photometric dispersion while fitting a transit model including instrumental systematic effects present in the time series.  

\subsubsection{MuSCAT}\label{muscat}
We observed an egress of TOI-530b on UT 2020 March 2 in $g$, $r$, and $z_s$ bands, using the multiband imager MuSCAT \citep{2015JATIS...1d5001N} mounted on the 188~cm telescope of National Astronomical Observatory of Japan at the Okayama Astro-Complex, Japan. MuSCAT has three CCD cameras, each having a pixel scale of $0.361''$ pixel$^{-1}$ and a field of view of $6.1' \times 6.1$. We acquired 321, 268, and 474 images with exposure times of 30, 30, and 20~s in $g$, $r$, and $z_s$ bands, respectively. The data were dark-subtracted and flat-field corrected in a standard manner. Aperture photometry was then performed on the reduced images using a custom pipeline \citep{2011PASJ...63..287F}. The radius of the photometric aperture was chosen to be 18 pixels ($6.5''$) for all bands so that the photometric dispersion was minimized.

\begin{table*}
    \centering
    \caption{Ground-based photometric follow-up observations for \tar}
    \begin{tabular}{ccccccccc}
        \hline\hline
        Telescope &Camera &Filter &Pixel Scale & Aperture Size (pixel) &Coverage      &Date &Duration (minutes) &Total exposures  \\\hline
        El Sauce (0.36 m) &STT 1603 &$R_{c}$ &1.47 &4 &Egress &2019 November 21 &183 &59\\\hline
        TCS (1.52 m) &MuSCAT2 &$g$ &0.44 &9.8 &Egress &	2020 January 4 &237 &305\\
        TCS (1.52 m) &MuSCAT2 &$r$ &0.44 &9.8 &Egress &	2020 January 4 &237 &456\\
        TCS (1.52 m) &MuSCAT2 &$i$ &0.44 &9.8 &Egress &	2020 January 4 &237 &456\\
        TCS (1.52 m) &MuSCAT2 &$z_{s}$ &0.44 &9.8 &Egress &	2020 January 4 &237 &238\\\hline
        NAOJ (1.88 m) &MuSCAT &$g$ &0.36 &18 &Egress &2020 March 2 &177 &321\\
        NAOJ (1.88 m) &MuSCAT &$r$ &0.36 &18 &Egress &2020 March 2 &177 &268\\
        NAOJ (1.88 m) &MuSCAT &$z_{s}$ &0.36 &18 &Egress &2020 March 2 &177 &474\\
         \hline
    \end{tabular}
    \label{po}
\end{table*}

\begin{figure*}
\centering
\includegraphics[width=\textwidth]{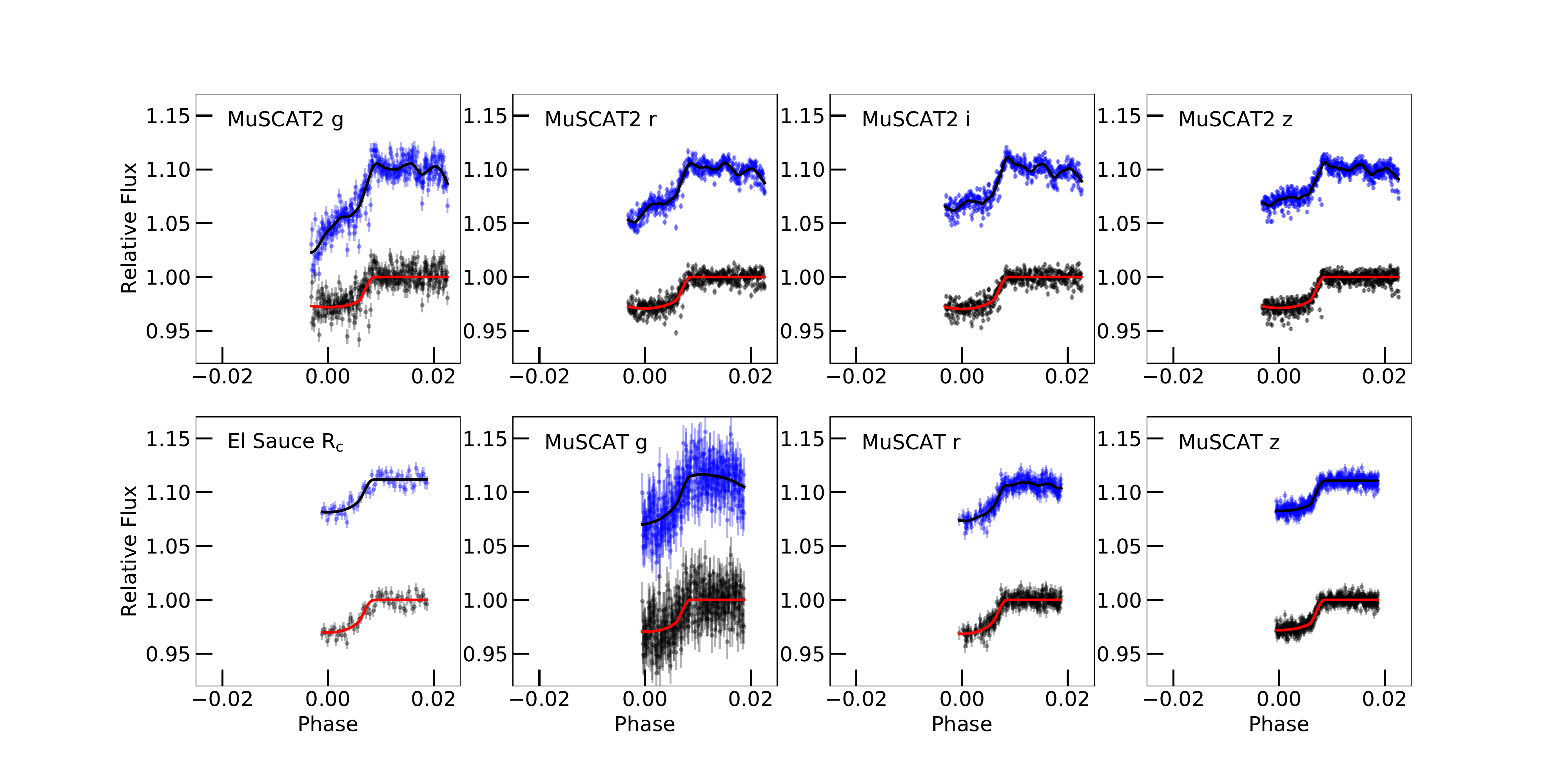}
\caption{Ground-based light curves for all available instrument. The blue dots are the raw data while the black solid line represents the best fit GP+transit model. The black dots are results after subtracting the GP model (i.e., detrended data). We use these detrended light curves in the final joint-fit (see Section \ref{joint}).} 
\label{ground_transit_detrend}
\end{figure*}

\subsection{Spectroscopic Observations}
\subsubsection{IRTF}
We observed TOI-530 on UT 2019 April 23 with the uSpeX spectrograph \citep{2003PASP..115..362R, 2004SPIE.5492.1498R} on the 3-m NASA Infrared Telescope Facility (IRTF). Our data was collected in the SXD mode using the $0\farcs3 \times 15\arcsec$ slit and covers a wavelength range of $0.7-2.55$ $\mu$m. The data was reduced using the \texttt{Spextool} pipeline \citep{2004PASP..116..362C}. After reducing, we RV-correct our spectrum using \texttt{tellrv} \citep{2014AJ....147...20N}, with which we estimate a systemic radial velocity of $-26 \pm 5$ km/s. By comparing our spectrum to those provided by the IRTF library \citep{Rayner2009}, we determine that our spectrum best matches that of a star of spectral type M0.5V. Lastly, we calculate the metallicity of TOI-530 following the relations defined in \cite{Mann2013} for cool dwarfs with spectral types between K7 and M5. In performing this calculation, we opted to only use the $Ks$-band spectrum, as \cite{2019AJ....158...87D} found $Ks$-band spectra to produce more reliable metallicities and suffer less telluric contamination than $H$-band spectra. Our analysis yield metallicities of [Fe/H] = $0.376 \pm 0.095$ and [M/H] = $0.218 \pm 0.092$.

\begin{figure}
\centering
\includegraphics[width=0.49\textwidth]{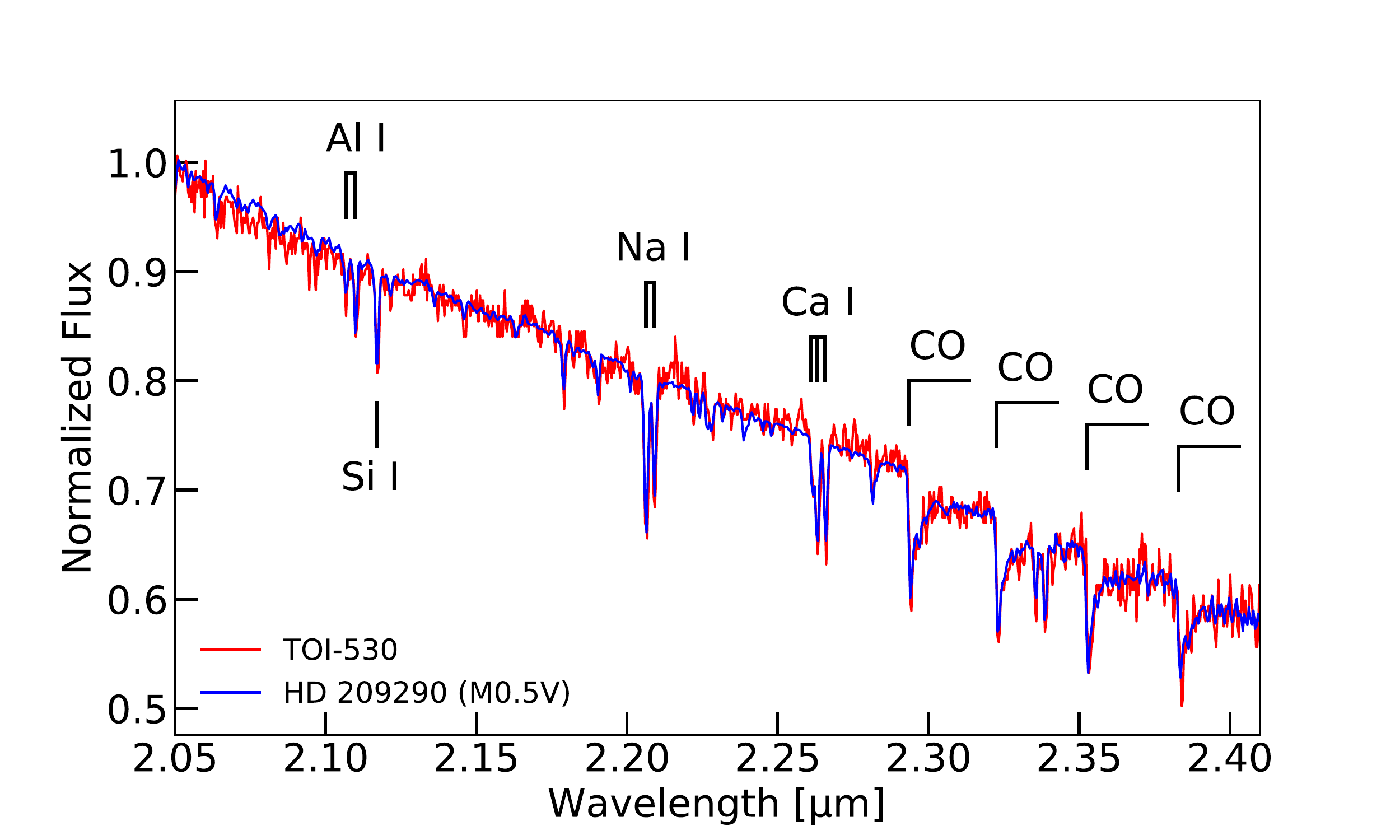}
\caption{Renormalized SpeX spectrum of \tar\ (red line) and the comparison spectrum (blue line) taken from the IRTF library \citep{Rayner2009}. The strong atomic features are marked based on the results from \citet{Cushing2005}. The NIR spectrum of \tar\ corresponds to a spectral type of M0.5V.} 
\label{IRTF}
\end{figure}

\subsubsection{CFHT/SPIRou}
We monitored \tar\ over 5 epochs between UT 2020 September 26 and UT 2020 October 5 using SPIRou (standing for SpectroPolarim\`etre InfraROUge), which is a new-generation high resolution ($64,000$) fiber-fed spectrograph with polarimetric and precision velocimetry capacities, installed at CFHT in 2018 \citep{Artigau2014,Donati2018}. It has a large bandwidth (from 0.95 to 2.5 $\mu$m) allowing the detection of several stellar lines in a single shot thus enhancing the precision of the measurement of the stellar radial velocity. For each night, we obtained three sequences, with 975s exposure time for each. The spectroscopic data were reduced using the standard data reduction pipeline (APERO, Cook et al. in prep), which performs the data calibration and corrects the telluric and night-sky emission \citep{Artigau2021}. For the Night-sky emission, it is corrected using a principal component analysis (PCA) model of OH emission constructed from a library of high-SNR sky observations \citep{Artigau2014PCA}. The telluric absorption is corrected using a PCA-based approach on residuals after fitting for a basic atmospheric transmission model (TAPAS, \citealt{Bertaux2014}).

We extracted the RVs of TOI-530 from the telluric-subtracted SPIRou spectra using \code{wobble} \citep{Bedell2019}. Briefly, \code{wobble} constructs a linear model to infer the stellar and time-varying telluric spectra without requiring any prior knowledge on them, while solving for the RV at each epoch. 

We only used the orders 29--37 (around 1490--1800~nm, all in the H band) to extract RVs for the following reasons. We dropped the orders 13--17, 23--28, 39--43, and 47--49 (around 1130--1230~nm, 1330--1490~nm, 1850--2010~nm, 2326--2510~nm, respectively), because the telluric absorption lines are too heavy. These orders with heavy telluric absorption are basically the wavelength regions in between the photometric bands (Y, J, H, K, and at the end of the K band). Furthermore, we dropped the orders 1--12, 18--22, 44--46 (around 965--1140~nm, 1225--1340~nm, and 2132--2290~nm, respectively), because their signal-to-noise ratios (SNRs) are too low (lower than $\sim$30). The low SNRs caused poor corrections of the telluric lines by the SPIRou pipeline, which manifests as significant residuals of telluric emission/absorption, as well as some abnormal features caused by improper telluric subtraction. In addition, the authors of \code{wobble} also cautioned regarding applying the code to spectral data with SNR less than 50 \citep[][e.g., in their example using the Barnard's star's data]{Bedell2019}. We extracted RVs from the low-SNR orders and saw little RV variations in these orders, which we believe is due to the fact that \code{wobble} struggles to recover any RV information from these low-SNR spectra with heavy telluric residuals. 

To pre-process the spectra, we first dropped 500 pixels at both edges of each order and masked out the occasional residual emission lines not fully subtracted by the SPIRou pipeline as follows: We calculated the 80th percentile of the flux (denoted as $q$) in any given order, labeled all pixels with flux larger than $3 \times q$ as the emission-line pixels, and masked out 5 pixels in total centered around them. Then we scaled the blaze function offered by the SPIRou pipeline to the flux level of the observed spectrum and calculated the flux minus the blaze function at each pixel. If the difference is more than half of the maximum of the blaze function of corresponding order, such pixels were considered as emission-line pixels, and 5 pixels around them were masked out. Then, we used the scaled blaze function to continuum-normalize the spectra. 

Next, we passed the natural log of the wavelength, the natural log of the flux, the estimated inverse variance of the flux (set as photon counts at each pixel, assuming Poisson noise on the flux), the time of the observations, the BERVs and the airmass values to \code{wobble}. We let \code{wobble} only infer the stellar the spectra to extract RVs because the SPIRou pipeline already divided out telluric absorption. When \code{wobble} infers the stellar spectrum, it needs optimized L1 and L2 regularization parameters for each orders. For simplicity, we set these regularization parameters to the default values in the \code{wobble} code, which are the same for all orders.

To validate our work, we divided each order into the left part and the right part so that the total photon counts of each part are equal. Then we used the same method to get RVs from each part respectively. Comparing the RVs from the left and the right parts of each order, we found that the differences are on par with the RV differences between the three observations taken on the same night (i.e., the intra-night RV variation as derived using the full order). The RV signals are basically consistent cross the nine orders we analyzed. This suggests that our results are unlikely to arise from random noise but instead are of real astrophysical origin. However, we found that the differences between the RVs reported from the left or the right parts of each order (typically 10--30 m/s) are significantly larger than the RV error bars reported by \code{wobble} (typically 1.6--1.9 m/s). Therefore, we calculated the standard deviation of the six RVs from the left and the right of each night (two RVs per observation, 3 observations per night) and used them as the more realistic estimates of the uncertainties of the RVs, which are what we present in Table~\ref{spirourv}.

\begin{table}
     \centering
     \caption{SPIRou RV measurements of \tar. Each observation took an exposure time of 975s. The RV offset here is arbitrary.}
     \begin{tabular}{ccc}
         \hline\hline
         BJD$_\mathrm{TDB}$       &RV\ (m~s$^{-1}$) &$\sigma_{\rm RV}$\ (m~s$^{-1}$) \\\hline
         2459119.08206 	&29356.43 	&20.14 \\ 
         2459119.09361 	&29372.09 	&20.14 \\ 
         2459119.10515 	&29389.86 	&20.14 \\
         2459120.06581 	&29486.04 	&12.55 \\
         2459120.07736 	&29497.80 	&12.55 \\
         2459120.08897 	&29500.96 	&12.55 \\
         2459123.06563 	&29342.45 	&14.13 \\
         2459123.07718 	&29358.37 	&14.13 \\
         2459123.08873 	&29364.25 	&14.13 \\
         2459127.07895 	&29461.84 	&20.86 \\
         2459127.09089 	&29495.09 	&20.86 \\
         2459127.10289 	&29503.15 	&20.86 \\
         2459128.06393 	&29353.01 	&31.13 \\
         2459128.07548 	&29374.36 	&31.13 \\
         2459128.08703 	&29445.57 	&31.13 \\
          \hline
     \end{tabular}
     \label{spirourv}
\end{table}

\subsection{High Angular Resolution Imaging}
If an exoplanet host star has a spatially close companion, that companion (bound or line of sight) can create a false-positive transit signal if it is, for example, an eclipsing binary (EB). For small stars and large planets, this is an especially important check to make, due to the paucity of giant planets orbiting M stars. ``Third-light” flux from the close companion star can lead to an underestimated planetary radius if not accounted for in the transit model \citep{Ciardi2015} and cause non-detections of small planets residing with the same exoplanetary system \citep{Lester2021}. Additionally, the discovery of close, bound companion stars, which exist in nearly one-half of FGK type stars \citep{Matson2018} and less so for M class stars, provides crucial information toward our understanding of exoplanetary formation, dynamics and evolution \citep{Howell2021}. Thus, to search for close-in bound companions unresolved in TESS or other ground-based follow-up observations, we obtained high-resolution imaging observations of TOI-530.

\subsubsection{Keck/NIRC2 Adaptive Optics Imaging}
We observed \tar\ with infrared high-resolution adaptive optics (AO) imaging at Keck Observatory \citep{Ciardi2015,Schlieder2021} on UT 2019 April 7. The observations were made with the NIRC2 instrument on Keck-II behind the natural guide star AO system. The standard 3-point dither pattern was used to avoid the left lower quadrant of the detector which is typically noisier than the other three quadrants. The dither pattern step size
was 3$''$ and it was repeated twice, with each dither offset from the previous one by 0.5$''$.

The observations were taken in the broad-band K ($\lambda_{o}=2.1956$ $\mu$m; $\Delta \lambda=0.336$ $\mu$m) with an integration time of 4 s per frame for a total on-source integration time of 36 s. The camera was in the narrow-angle mode with a full field of view of 10$''$ and a pixel scale of approximately 0.009942$''$ per pixel. The Keck AO observations show no additional stellar companions were detected to within a resolution $\sim0.056''$ FWHM. The sensitivities of the final combined AO image were determined by injecting simulated sources azimuthally around the primary target every $20^{\circ}$ at radial separations of integer multiples of the FWHM of the central source \citep{Furlan2017}. The brightness of each injected source was scaled until standard aperture photometry detected it with $5\sigma$ significance. The resulting brightness of the injected sources relative to the target \tar\ was regarded as the contrast limits at that injection location. The final $5\sigma$ limit at each separation was determined from the average of all of the determined limits at that separation while the uncertainty was given by the RMS dispersion of the results for different azimuthal slices at a given radial distance. We show the 2$\mu$m sensitivity curve in the left panel of Figure \ref{imaging} along with an inset image zoomed to primary target, which shows no other companion stars. 

\subsubsection{Gemini-North Speckle Imaging}
\begin{figure*}
\centering
\includegraphics[width=\textwidth]{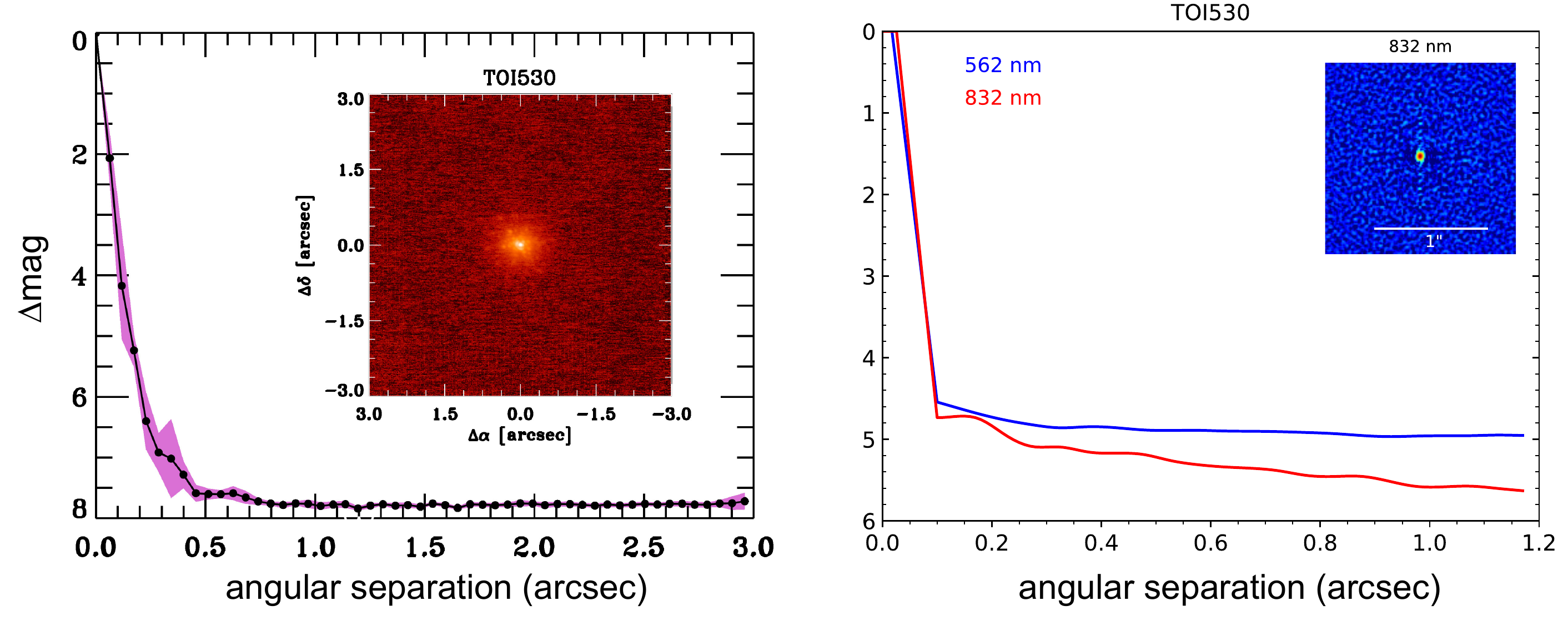}
\caption{{\it Left panel}: NIRC2 AO image (inset) and $\rm K_{s}$-band contrast curve for \tar. The black line is the $5\sigma$ sensitivity limit. The shaded purple region represents the azimuthal dispersion ($1\sigma$) of the contrast determinations. {\it Right panel}: The $5\sigma$ ‘Alopeke speckle imaging contrast curves in both filters as a function of the angular separation out to 1.2 arcsec, the end of speckle coherence. The inset shows the reconstructed 832 nm image with a 1 arcsec scale bar. The star, TOI-530, was found to have no close companions to within the contrast levels achieved. } 
\label{imaging}
\end{figure*}

TOI-530 was observed on 2020 February 17 UT using the ‘Alopeke speckle instrument on the Gemini North 8-m telescope\footnote {\url{https://www.gemini.edu/sciops/instruments/alopeke-zorro/}}.  ‘Alopeke provides simultaneous speckle imaging in two bands (562nm and 832 nm) with output data products including a reconstructed image with robust contrast limits on companion detections (e.g., \citealt{Howell2016}). Ten sets of $1000\times0.06$ sec exposures were collected and subjected to Fourier analysis in our standard reduction pipeline \citep{Howell2011}. The right panel of Figure \ref{imaging} shows our final contrast curves and the 832 nm reconstructed speckle image. We find that TOI-530 is a single star with no companion brighter than 5-6 magnitudes below that of the target star (earlier than $\sim$ M4.5V) from the diffraction limit (20 mas) out to 1.2$''$. At the distance of TOI-530 (d=148 pc) these angular limits correspond to spatial limits of 3 to 178 au.

\section{Stellar Characterization}\label{stellar_properties}

We first use 2MASS observed $m_{K}$ and the parallax from \gaia\ EDR3 to calculate the absolute magnitude, of which we obtain $M_{K}= 5.42 \pm 0.13$ mag. We then estimate the stellar radius following the polynomial relation between $R_{\ast}$ and $M_{K}$ derived by \cite{Mann2015}, and we find $R_{\ast}=0.54\pm0.02\ R_{\odot}$, assuming a typical uncertainty of 3\% (see Table 1 in \citealt{Mann2015}). For comparison, we also estimate the stellar radius $R_{\ast}=0.55\pm0.03 R_{\odot}$ based on the angular diameter relation in \cite{Boyajian2014}, consistent with our previous estimate within $1\sigma$. 

Using the empirical polynomial relation between bolometric correction ${\rm BC}_{K}$ and $V-J$ in \cite{Mann2015}, we find ${\rm BC}_{K}$ to be $2.60\pm0.13$ mag. Thus, we derive a bolometric magnitude $M_{\rm bol}=8.02\pm 0.13$ mag, leading to a bolometric luminosity of $L_{\ast}=0.049\pm0.005\ L_{\odot}$. To estimate the stellar effective temperature of \tar, we first take use of the Stefan-Boltzmann law. Coupled with the aforementioned stellar radius and bolometric luminosity we derived, we get $T_{\rm eff}=3666\pm146$ K. As an independent check, we then obtain $T_{\rm eff}$ following the empirical relation reported by \cite{Mann2015} and we find $T_{\rm eff}=3650\pm100$ K. Both estimations agree well with the result $T_{\rm eff}=3663\pm124$ K from \cite{Pecaut2013}. 

Finally, we evaluate that \tar\ has a mass of $M_{\ast}=0.53\pm0.01\ M_{\odot}$ using Equation 2 in \cite{Mann2019} according to the $M_{\ast}$-$M_{K}$ relation. This is consistent with the value $M_\ast = 0.52 \pm 0.03\ M_\odot$ given by the eclipsing-binary based empirical relation of \citet{Torres2010}. 

\begin{figure}
    \centering
    \includegraphics[width=\linewidth,trim=10 5 20 15,clip]{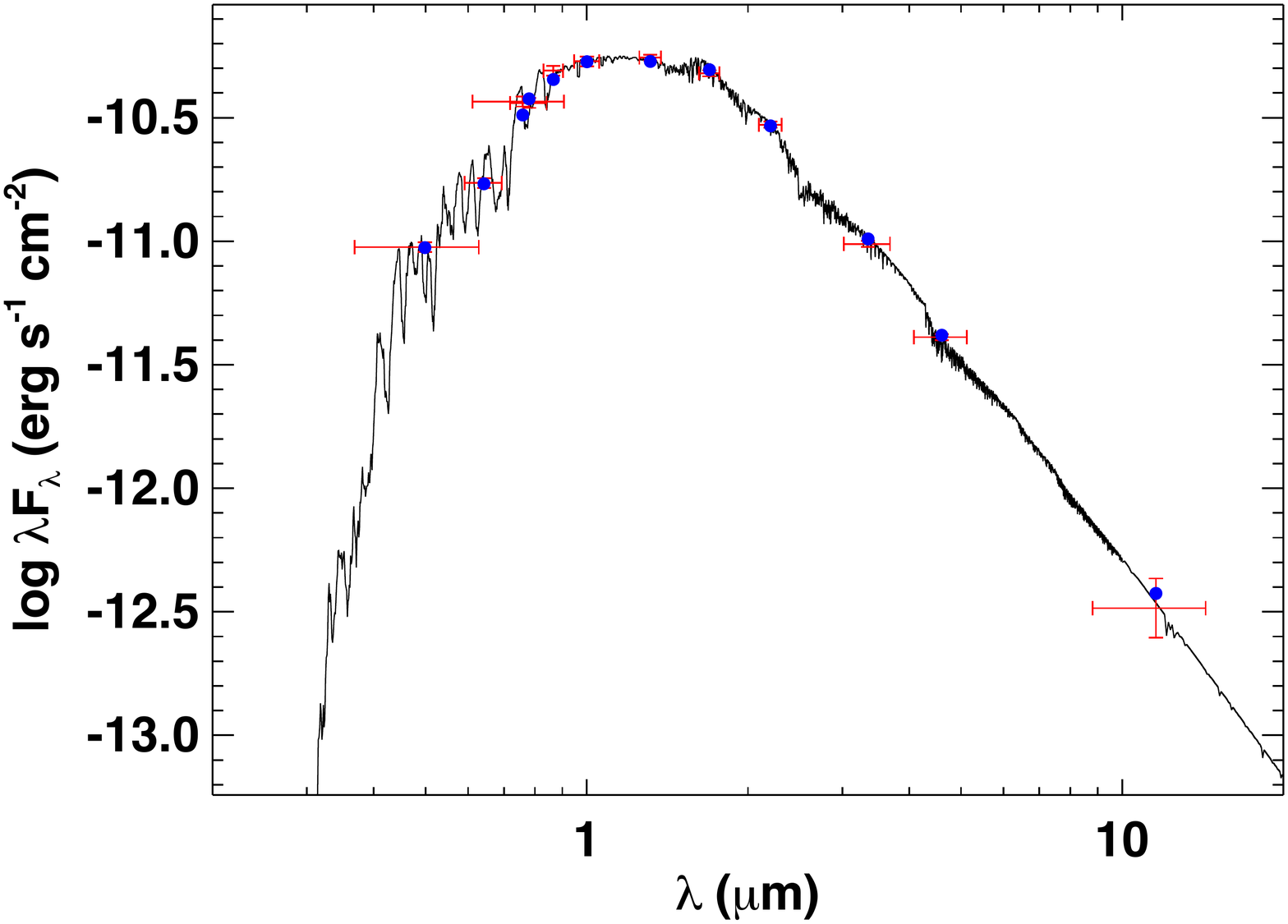}
    \caption{The best SED fit for \tar. Red symbols represent the observed photometric measurements, where the horizontal bars represent the effective width of the passband. Blue symbols are the model fluxes from the best-fit NextGen atmosphere model (black). 
\label{fig:sed}}
\end{figure}

As an independent check, we carry out an analysis of the broadband Spectral Energy Distribution (SED) together with the {\it Gaia\/} EDR3 parallax in order to determine an independent, empirical measurement of the stellar radius, following the procedures described in \citet{Stassun2016}, \citet{Stassun2017}, and \citet{Stassun2018}. We pull the $JHK_S$ magnitudes from {\it 2MASS} \citep{Cutri2003,skrutskie2006}, the W1--W3 magnitudes from {\it WISE} \citep{wright2010},  the $grizy$ magnitudes from Pan-STARRS \citep{Magnier2013}, and three \gaia\ magnitudes $G, G_{\rm BP}, G_{\rm RP}$ \citep{GaiaEDR3}. Together, the available photometry spans the full stellar SED over the wavelength range 0.4\,--\,10~$\mu$m (see Figure~\ref{fig:sed}). 

We perform a fit using NextGen stellar atmosphere models, with the $T_{\rm eff}$, $\log g$, and [Fe/H] taken from the spectroscopic analysis. The remaining parameter is the extinction ($A_V$), which we limit to the full line-of-sight extinction from the dust maps of \citet{schlegel1998}. The resulting fit is shown in Figure.~\ref{fig:sed} with a reduced $\chi^2$ of 1.6 and best-fit extinction of $A_V = 0.00^{+0.03}_{-0.00}$. Integrating the model SED gives the bolometric flux at Earth of \mbox{$F_{\rm bol} =  7.009 \pm 0.081 \times 10^{-11}$ erg~s$^{-1}$~cm$^{-2}$}. Taking the $F_{\rm bol}$ and $T_{\rm eff}$ together with the {\it Gaia\/} parallax, with no adjustment for systematic parallax offset \citep[see, e.g.,][]{Stassun2021}, gives the stellar radius as $R_\ast = 0.547 \pm 0.030\ R_\odot$. 

Combining all the results above, we adopt the weighted-mean values of effective temperature $T_{\rm eff}$, stellar radius $R_{\ast}$ and stellar mass $M_{\ast}$ as listed in Table \ref{starparam}. 

To identify the Galactic population membership of \tar, we first calculate the three-dimensional space motion with respect to the LSR based on \cite{Johnson1987}. We adopt the astrometric values ($\varpi$, $\mu_{\alpha}$, $\mu_{\delta}$) from \gaia\ EDR3 and the spectroscopically determined systemic RV from the SpeX spectrum, and we find $U_{\rm LSR}=48.99\pm4.59$ km s$^{-1}$, $V_{\rm LSR}=-20.10\pm1.92$ km s$^{-1}$, $W_{\rm LSR}=-6.73\pm0.56$ km s$^{-1}$. Following the procedure described in \cite{Bensby2003}, we compute the relative probability $P_{\rm thick}/P_{\rm thin}=0.02$ of \tar\ to be in the thick and thin disks by taking use of the recent kinematic values from \cite{Bensby2014}, indicating that \tar\ belongs to the thin-disk population. We further integrate the stellar orbit with the ``MWPotential2014'' Galactic potential using \code{galpy} \citep{Bovy2015} following \cite{Gan2020}, and we estimate that the maximal height $Z_{\rm max}$ of \tar\ above the Galactic plane is about $109$ pc, which agrees with our thin-disk conclusion. 

We finally perform a frequency analysis on the \tess\ PDCSAP photometry after masking the known in-transit data using the generalized Lomb-Scargle periodogram \citep{Zechmeister2009} to look for stellar activity signals. We find a peak at around $9.4$ d in the \tess\ Sector 6 data, which may be attributed to stellar rotation. However, this periodic signal is not significant in the generalized Lomb-Scargle periodogram of the \tess\ photometry taken in the extended mission. We further analyze the ground-based long-term photometry from the Zwicky Transient Facility (ZTF; \citealt{Masci2019}). ZTF took a total of 273 exposures for \tar, which spanned 1036 d. We clip outliers above the $3\sigma$ level and 242 measurements are left. However, we find that the 9.4 d signal does not show up in the corresponding generalized Lomb-Scargle periodogram, either. Additionally, \citealt{Newton2018} shows a typical rotational period of $\sim40$ d for a $0.5\ M_{\odot}$ star. We thus conclude that the $9.4$ d signal is probably not associated with stellar rotation. Future \tess\ data to be obtained will allow better identification of the correct rotation period of this target. 


\begin{figure}
\centering
\includegraphics[width=0.49\textwidth]{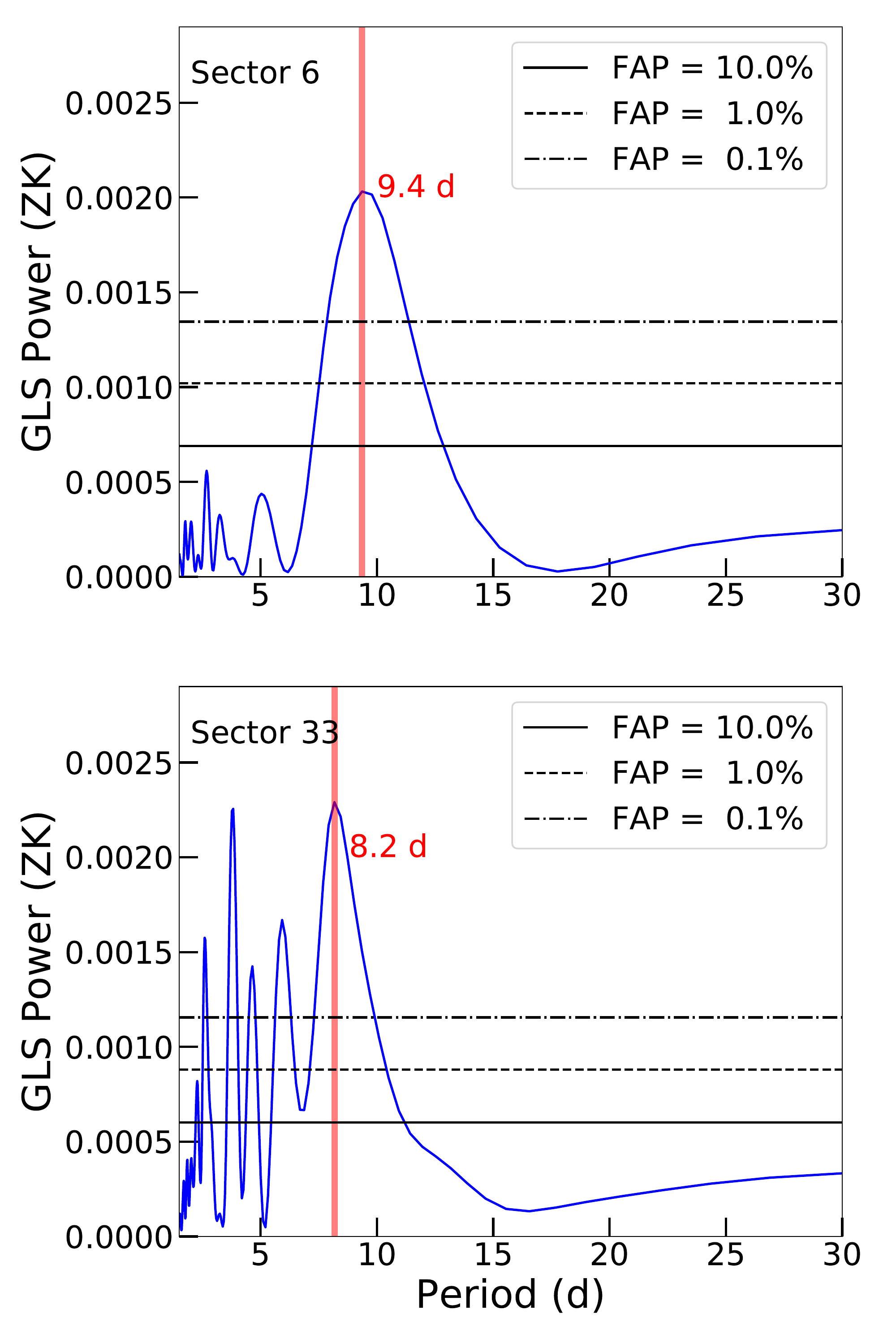}
\caption{Generalized Lomb-Scargle periodograms of the TESS PDCSAP photometry from two sectors. The theoretical FAP levels of 10, 1, and 0.1 percent are marked as horizontal solid, dashed, and dot–dashed lines. The vertical red lines mark the maximum peaks of the periodograms.} 
\label{TESS_GLS}
\end{figure}



\begin{table}\label{stellarinfor}
    \caption{Basic information of \tar}
    \begin{tabular}{lll}
        \hline\hline
        Parameter       &Value       \\\hline
        \it{Main identifiers}                    \\
         TOI                     &$530$         \\
         TIC                     &$387690507$   \\
         \gaia\ ID            &$3353218995355814656$ \\
         \it{Equatorial Coordinates} \\
         $\rm R.A.\ (J2015.5)$                    &06:53:39.08 \\
         $\rm DEC.\ (J2015.5)$                    &12:52:53.68    \\
         \it{Photometric properties}\\
         $\tess$\ (mag)           &$13.5287\pm0.0076$   &$\rm TIC\ V8^{[1]}$     \\
         $\gaia$\ (mag)           &$14.6217\pm0.0006$   &\gaia\ EDR3$^{[2]}$   \\
         \gaia\ BP\ (mag)           &$15.814\pm0.004$   &\gaia\ EDR3   \\
         \gaia\ RP\ (mag)           &$13.538\pm0.002$   &\gaia\ EDR3   \\
         $B$\ (mag)                    &$16.708\pm0.044$         &APASS \\
         $V$\ (mag)                    &$15.403\pm0.136$          &APASS \\
         $J$\ (mag)                    &$12.112\pm0.023$   &2MASS \\
         $H$\ (mag)                    &$11.468\pm0.030$   &2MASS \\
         $K$\ (mag)                    &$11.238\pm0.020$    &2MASS \\
         \wise1 (mag)                   &$11.124\pm0.023$   &\wise \\
         \wise2 (mag)                   &$11.087\pm0.020$   &\wise \\
         \wise3 (mag)                   &$10.907\pm0.139$   &\wise \\
         \wise4 (mag)                   &$8.735\pm0.429$   &\wise \\
         \it{Astrometric properties}\\
         $\varpi$ (mas)              &$6.77\pm0.02$  &\gaia\ EDR3  \\
         $\mu_{\rm \alpha}\ ({\rm mas~yr^{-1}})$     &$13.62\pm0.03$   &\gaia\ EDR3   \\
         $\mu_{\rm \delta}\ ({\rm mas~yr^{-1}})$     &$-62.52\pm0.02$   &\gaia\ EDR3  \\
         RV\ (km~s$^{-1}$)                          &$-25.93\pm2.00$ &This work  \\
         \it{Derived parameters} \\
         Distance (pc)                &$147.7\pm0.6$  &This work     \\
         $U_{\rm LSR}$ (km~s$^{-1}$)       &$48.99\pm4.59$     &This work\\
         $V_{\rm LSR}$ (km~s$^{-1}$)       &$-20.10\pm1.92$     &This work\\
         $W_{\rm LSR}$ (km~s$^{-1}$)       &$-6.73\pm0.56$     &This work\\
         $M_{\ast}\ (M_{\odot})$ &$0.53\pm 0.02$ &This work       \\
         $R_{\ast}\ (R_{\odot})$ &$0.54\pm 0.03$ &This work       \\
         $\rho_\ast\ ({\rm g~cm^{-3}})$ &$4.74\pm 1.11$ &This work \\
         $\log g_{\ast}\ ({\rm cgs})$       &$4.70\pm 0.03$  &This work        \\
         $L_{\ast}\ (L_{\odot})$ &$0.049\pm0.005$  &This work    \\
         $T_{\rm eff}\ ({\rm K})$           &$3659\pm 120$  &This work       \\
         $\rm [Fe/H]$  &$0.376\pm 0.095$ &This work \\
         $\rm [M/H]$  &$0.218\pm0.092$ &This work\\
         \hline\hline 
    \end{tabular}
    \begin{tablenotes}
    \item[1]  [1]\ \cite{Stassun2017tic,Stassun2019tic} 
    \item[2]  [2]\ \cite{GaiaEDR3}
    \end{tablenotes}
    \label{starparam}
\end{table}

\section{Analysis and results}\label{analysis}
\subsection{Photometric Analysis}\label{transit}
\subsubsection{\tess\ only}\label{tess_only}
We first model the detrended \tess\ only photometry by utilizing the \code{juliet} package \citep{juliet}, which employs \code{batman} to build the transit model \citep{Kreidberg2015}. Dynamic nested sampling is applied in \code{juliet} to determine the posterior estimates of system parameters using the publicly available package \code{dynesty} \citep{Higson2019,Speagle2019}.

We set uninformative uniform priors on both the transit epoch ($T_{0}$) and the orbital period ($P_{b}$), centered on the optimized value obtained from the TLS analysis. Following the approach described in \cite{Espinoza2018}, instead of directly fitting for the radius ratio ($p=R_{p}/R_{\ast}$) and the impact parameter ($b=a/R_{\ast}\cos i$), we apply the new parametrizations $r_{1}$ and $r_{2}$ to sample points, for which we impose uniform priors between 0 and 1. This new parametrization allows us to only sample physically meaningful values of a transiting system with $0 < b < 1 + p$, which reduces the computational cost. We adopt a quadratic limb-darkening law for the \tess\ photometry, where we place a uniform prior on both coefficients ($q_{1}$ and $q_{2}$, \citealt{Kipping2013}). Since photometric-only data weakly constrain the orbital eccentricity, we fix $e$ at zero and include a non-informative log-uniform prior on stellar density. We fit an extra flux jitter term to account for additional systematics. As the \tess\ PDCSAP light curve has already been corrected for the light dilution, we fix the dilution factors $D$ to 1. Table \ref{tess_only_fit_priors} summarizes the prior settings we adopt as well as the best-fit value of each parameter. We then rerun the photometry-only fit with free $e$ and $w$ to examine potential evidences of eccentricity by comparing the Bayesian model log-evidence ($\ln Z$) difference between the circular and eccentric orbit models calculated using the \code{dynesty} package. Generally, we consider a model is strongly favored than another if $\Delta \ln Z>5$ \citep{Trotta2008}. We find that the circular orbit model is slightly preferred with a Bayesian evidence improvement of $\Delta \ln Z=\ln Z_{\rm Circular}-\ln Z_{\rm Keplerian}=2.8$. We thus conclude that there is no evidence of orbital eccentricity in the \tess\ time-series data. We use the posteriors from the circular orbit fit as a prior to detrend all ground-based photometric data (see next Section).

\subsubsection{Ground-based photometric data}\label{joint_fit_TESS_ground}
Since all of the eight ground light curves only covered partial transits, the way of detrending generally correlates with the final modeling results. Therefore, we decide to independently detrend all ground photometry in a uniform way using Gaussian processes. As there are no obvious quasi-periodic oscillations existing in data from different facilities, we choose the Mat\'{e}rn-3/2 kernel, formulated as:
\begin{equation}
    k_{i,j}(\tau) = \sigma^{2}\left(1+\frac{\sqrt{3}\tau}{\rho}\right){\rm exp}\left(\frac{\sqrt{3}\tau}{\rho}\right),
\end{equation}
where $\tau$ is the time-lag, and $\sigma$ and $\rho$ are the covariance amplitude and the correlation timescale of the GP, respectively. Taking the posteriors from the previous \tess\ only fit into account, we put a constraint on the priors to optimize the sampling and reduce the computational time cost. We list our priors in Table \ref{ground_only_fit_priors} and show the raw and detrended ground light curves in Figure \ref{ground_transit_detrend}.


\begin{figure}
\centering
\includegraphics[width=0.49\textwidth]{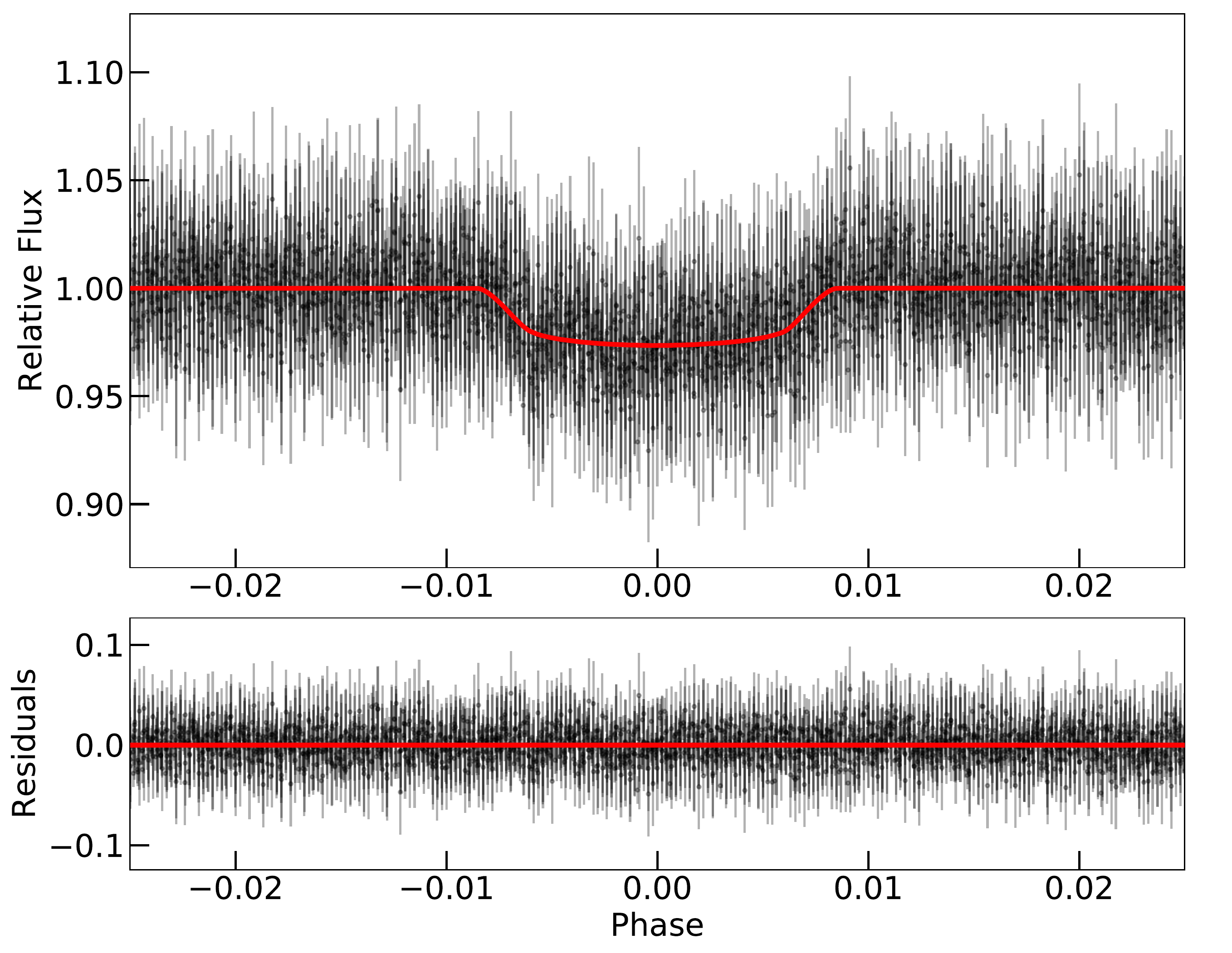}
\caption{{\it Top panel}: Phase-folded TESS photometry of \tar. The red solid line represents the median posterior model. {\it Bottom panel}: The residuals of the TESS data after subtracting the best-fit transit model. } 
\label{TESS_transit}
\end{figure}


\begin{figure}
\centering
\includegraphics[width=0.49\textwidth]{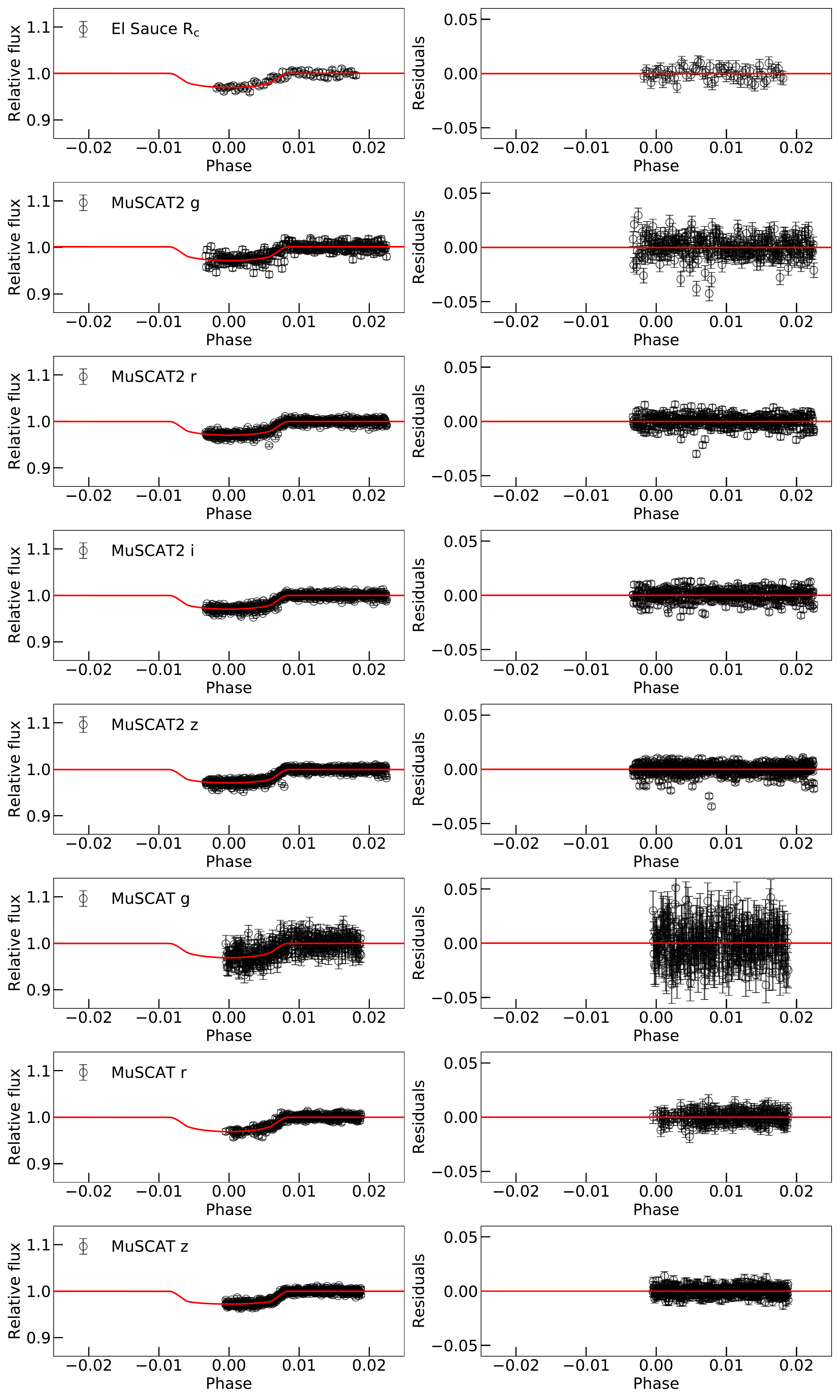}
\caption{Left panels show unbinned phase-folded follow-up transit light curves of \tar. The instrument and observational band information is presented at the top left of each panel. Our best-fit models are shown as red solid lines. The residuals are shown in the right panels.} 
\label{ground_transit}
\end{figure}

\subsection{RV-only modeling}\label{rv}
We carry out a preliminary RV-only fit using \code{juliet}, which utilizes the \code{radvel} package to build the Keplerian model \citep{Fulton2018}. In order to reduce the potential errors induced by the orbital period and timing, we fix $P_{b}$ and $T_{0,b}$ at the best-fit transit ephemeris derived from the previous \tess\ only fit. Due to the limited number of RV points and our previous insignificant detection of eccentricity (see Section \ref{tess_only}), we fit a circular orbit model with $e$ fixed at zero. Since our RV observations only have a short time span, we do not take the RV slope and quadratic term ($\dot{\gamma}$ and $\ddot{\gamma}$) into consideration in the RV modeling, and we simply fix them at 0. Thus the remaining degrees of freedom are the RV semi-amplitude $K_{b}$, the systemic velocity $\mu$ and the extra jitter term $\sigma$, which is used to account for the additional white noise. We adopt wide uniform priors on $K_{b}$ and $\mu$ but a log-uniform prior on $\sigma$. Our model reveals that the SPIRou RVs have a semi-amplitude of $K=67.2\pm15.1$ m/s. Table \ref{rvonly_priors} provides our prior settings and the median value of the posterior of each parameter along with their $1\sigma$ confidence interval.  

We then construct a flat RV model to test the robustness of our RV detection above. Compared with the flat model, we find that our circular orbit model has a $\ln Z$ improvement of $\Delta \ln Z=\ln Z_{\rm Circular}-\ln Z_{\rm Flat}=4.8$, supporting a significant RV detection.

\subsection{Joint RV and transit analysis}\label{joint}

\begin{figure*}
\centering
\includegraphics[width=\textwidth]{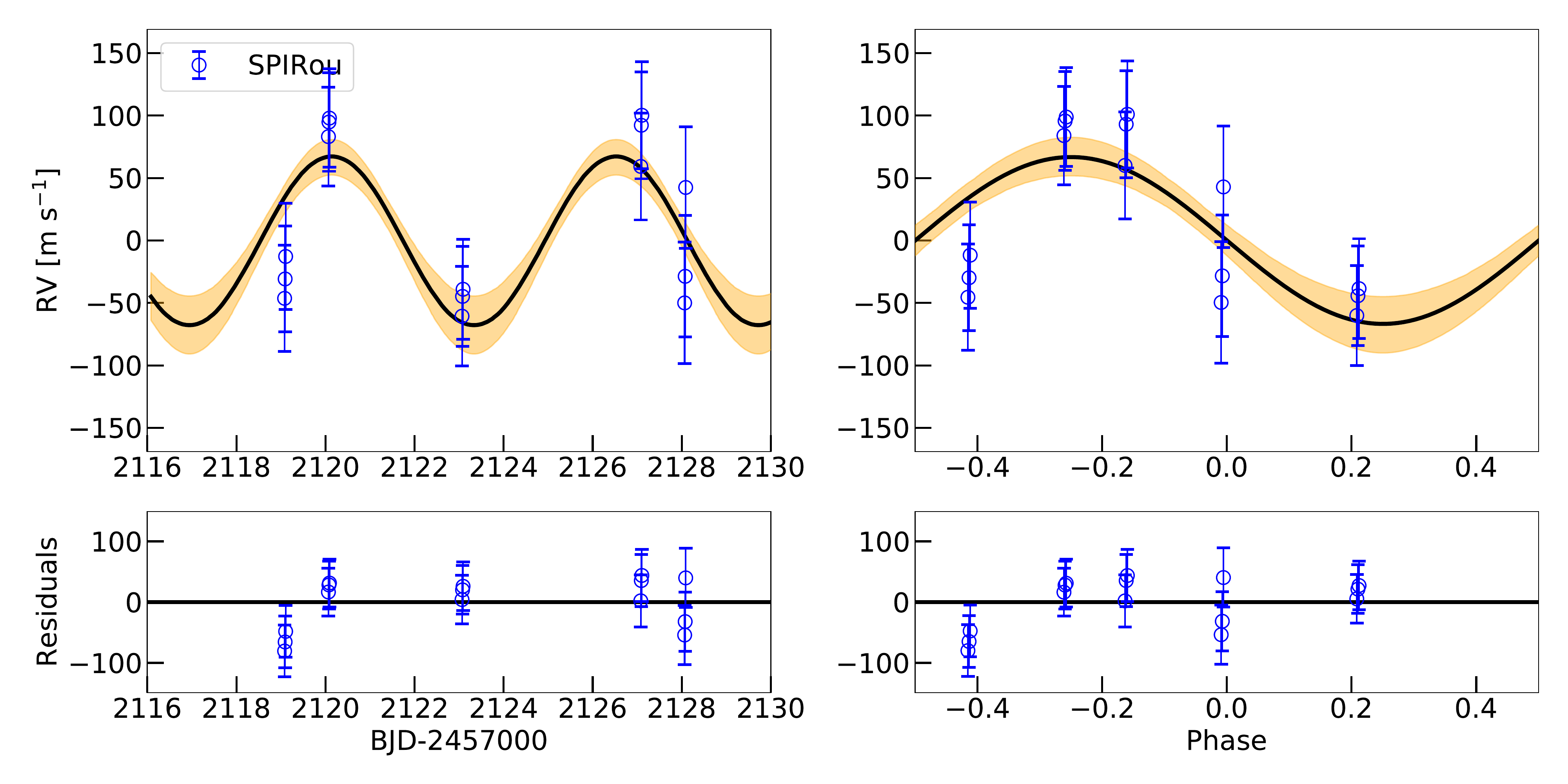}
\caption{{\it Left panel}: The systemic velocity-subtracted SPIRou RVs of \tar\ as a function of time along with the best-fit circular orbit model from the photometry+RV joint analysis shown as a black solid line. The error bars are the quadrature sum of the instrument jitter term and the measurement uncertainties for all RVs. The orange shaded region represents the $1\sigma$ confidence interval of the model. {\it Right panel}: The corresponding phased-folded SPIRou RV data. Residuals are plotted below.} 
\label{SPIROU}
\end{figure*}

In order to obtain precise transit ephemeris and physical parameters, we finally jointly model the detrended \tess\ photometry and all ground-based re-processed light curves together with the SPIRou RVs. We adopt the identical priors on planetary and \tess\ photometry parameters as in Section \ref{tess_only}. While for the ground photometric data, we choose the linear law to parameterize the limb-darkening effect and put a Gaussian prior on the theoretical estimate derived from the \code{LDTK} package with a width of 0.1 \citep{Husser2013,Parviainen2015}. Similarly, we also fit an extra flux jitter term for each ground instrument to account for additional white noise. As there are less contamination in the ground data, we fix all dilution factors $D$ to 1. For the SPIRou radial velocities, we adopt the same priors as the circular orbit model in Section \ref{rv}. We find the \tar b has a mass of $0.4\pm0.1\ M_J$ with a radius of $0.83\pm0.05\ R_{J}$, which is the typical size of a giant planet without much inflation. We show the phase-folded light curves along with the best-fit models in Figures \ref{TESS_transit} and \ref{ground_transit}. Figure \ref{SPIROU} shows the SPIRou data and the best-fit RV model.  Table \ref{tranpriors} summarizes the priors we set in the final joint fit as well as the best-fit value of each parameter. We list the final derived physical parameters in Table \ref{physical_parameter}. 

Since there are a total of 5 nearby stars of \tar\ with $T{\rm mag}<15.5$ located within $1'$ and the light from the brightest star among them (Gaia DR2 3353218784898973312, $T{\rm mag}=11.0$; star 5 in the right panel of Figure \ref{fov}) is expected to have a significant contribution of the contamination flux in the photometric aperture due to the large \tess\ pixel scale ($21''$/pixel), we rerun the joint fit to examine whether additional dilution correction is needed. We set a Gaussian prior on the \tess\ dilution factor $D_{\rm TESS}$, centered at 1 with a $1\sigma$ width of 0.1, and keep the left prior settings the same as above. We obtain $D_{\rm TESS}=0.97\pm0.03$ and a radius ratio of $R_{p}/R_{\ast}=0.156\pm0.001$, consistent with the result without considering light correction. 

\begin{table*}
    \caption{Prior settings and the best-fit values along with the 68\% credibility intervals in the final joint fit for \tar. $\mathcal{N}$($\mu\ ,\ \sigma^{2}$) means a normal prior with mean $\mu$ and standard deviation $\sigma$. $\mathcal{U}$(a\ , \ b) stands for a uniform prior ranging from a to b. $\mathcal{J}$(a\ , \ b) stands for a Jeffrey's prior ranging from a to b.}
    \begin{tabular}{lccr}
        \hline\hline
        Parameter       &Prior &Best-fit    &Description\\\hline
        \it{Planetary parameters}\\
        $P_{b}$ (days)  &$\mathcal{U}$ ($6.2$\ ,\ $6.6$)  &$6.387597^{+0.000019}_{-0.000018}$
        &Orbital period of \tar b.\\
        $T_{0,b}$ (BJD-2457000)  &$\mathcal{U}$ ($1468$\ ,\ $1472$)   &$1470.1998^{+0.0016}_{-0.0017}$
        &Mid-transit time of \tar b.\\
        $r_{1,b}$  &$\mathcal{U}$ (0\ ,\ 1)   &$0.553^{+0.056}_{-0.074}$
        &Parametrisation for {\it p} and {\it b}.\\
        $r_{2,b}$  &$\mathcal{U}$ (0\ ,\ 1)   &$0.155^{+0.002}_{-0.002}$
        &Parametrisation for {\it p} and {\it b}.\\
        $e_{b}$                     &0  &Fixed  &Orbital eccentricity of \tar b.\\
        $\omega_{b}$ (deg)          &90 &Fixed  &Argument of periapsis of \tar b.\\
        \it{\tess\ photometry parameters}\\
        $D_{\rm TESS}$  &Fixed    &$1$      &\tess\ photometric dilution factor.\\
        $M_{\rm TESS}$  &$\mathcal{N}$ (0\ ,\ $0.1^{2}$)   &$-0.00002^{+0.00009}_{-0.00009}$      &Mean out-of-transit flux of \tess\ photometry.\\
        $\sigma_{\rm TESS}$ (ppm) &$\mathcal{J}$ ($10^{-6}$\ ,\ $10^{6}$)  &$0.02^{+10.40}_{-0.01}$
          &\tess\ additive photometric jitter term.\\
        $q_{1}$       &$\mathcal{U}$ (0\ ,\ 1)           &$0.16^{+0.14}_{-0.09}$   &Quadratic limb darkening coefficient.\\
        $q_{2}$       &$\mathcal{U}$ (0\ ,\ 1)                 &$0.46^{+0.33}_{-0.30}$  &Quadratic limb darkening coefficient.\\
        
        \it{El Sauce photometry parameters}\\
        $D_{\rm el}$  &Fixed    &$1$      &El Sauce photometric dilution factor.\\
        $M_{\rm el}$  &$\mathcal{N}$ (0\ ,\ $0.1^{2}$)    &$-0.0004^{+0.0008}_{-0.0008}$
        &Mean out-of-transit flux of El Sauce photometry.\\
        $\sigma_{\rm el}$ (ppm) &$\mathcal{J}$ ($0.1$\ ,\ $10^{5}$)  &$17.3^{+483.7}_{-16.6}$
          &El Sauce additive photometric jitter term.\\
        $q_{\rm el}$   &$\mathcal{N}$ ($0.66$\ ,\ $0.1^{2}$)                &$0.74^{+0.07}_{-0.08}$     &Linear limb darkening coefficient.\\
        
        \it{MUSCAT2 photometry parameters}\\
        $D_{\rm MUSCAT2,g}$   &Fixed  &$1$
          &MUSCAT2 $g$ band photometric dilution factor.\\
        $M_{\rm MUSCAT2,g}$  &$\mathcal{N}$ (0\ ,\ $0.1^{2}$)   &$-0.0016^{+0.0006}_{-0.0005}$
          &Mean out-of-transit flux of MUSCAT2 $g$ band photometry.\\
        $\sigma_{\rm MUSCAT2,g}$ (ppm) &$\mathcal{J}$ ($0.1$\ ,\ $10^{5}$)  &$6868.9^{+517.3}_{-515.7}$
      &MUSCAT2 $g$ band additive photometric jitter term.\\
        $q_{\rm MUSCAT2,g}$      &$\mathcal{N}$ ($0.79$\ ,\ $0.1^{2}$)            &$0.67^{+0.07}_{-0.07}$        &Linear limb darkening coefficient.\\
        
        $D_{\rm MUSCAT2,r}$  &Fixed    &$1$
          &MUSCAT2 $r$ band photometric dilution factor.\\
        $M_{\rm MUSCAT2,r}$  &$\mathcal{N}$ (0\ ,\ $0.1^{2}$)   &$0.0001^{+0.0003}_{-0.0003}$
       &Mean out-of-transit flux of MUSCAT2 $r$ band photometry.\\
        $\sigma_{\rm MUSCAT2,r}$ (ppm) &$\mathcal{J}$ ($0.1$\ ,\ $10^{5}$)  &$4667.1^{+219.3}_{-203.6}$
       &MUSCAT2 $r$ band additive photometric jitter term.\\
        $q_{\rm MUSCAT2,r}$     &$\mathcal{N}$ ($0.73$\ ,\ $0.1^{2}$)             &$0.66^{+0.05}_{-0.05}$   &Linear limb darkening coefficient.\\
        
        $D_{\rm MUSCAT2,i}$  &Fixed     &$1$
          &MUSCAT2 $i$ band photometric dilution factor.\\
        $M_{\rm MUSCAT2,i}$  &$\mathcal{N}$ (0\ ,\ $0.1^{2}$)   &$0.0002^{+0.0003}_{-0.0003}$
          &Mean out-of-transit flux of MUSCAT2 $i$ band photometry.\\
        $\sigma_{\rm MUSCAT2,i}$ (ppm) &$\mathcal{J}$ ($0.1$\ ,\ $10^{5}$)  &$5026.9^{+207.0}_{-193.0}$
          &MUSCAT2 $i$ band additive photometric jitter term.\\
        $q_{\rm MUSCAT2,i}$     &$\mathcal{N}$ ($0.54$\ ,\ $0.1^{2}$)             &$0.61^{+0.05}_{-0.05}$   &Linear limb darkening coefficient.\\
        
        $D_{\rm MUSCAT2,z}$   &Fixed  &$1$
          &MUSCAT2 $z$ band photometric dilution factor.\\
        $M_{\rm MUSCAT2,z}$  &$\mathcal{N}$ (0\ ,\ $0.1^{2}$)  &$0.0006^{+0.0002}_{-0.0002}$
          &Mean out-of-transit flux of MUSCAT2 $z$ band photometry.\\
        $\sigma_{\rm MUSCAT2,z}$ (ppm) &$\mathcal{J}$ ($0.1$\ ,\ $10^{5}$)  &$4504.2^{+144.9}_{-140.6}$
          &MUSCAT2 $z$ band additive photometric jitter term.\\
        $q_{\rm MUSCAT2,z}$      &$\mathcal{N}$ ($0.44$\ ,\ $0.1^{2}$)             &$0.58^{+0.05}_{-0.05}$  &Linear limb darkening coefficient.\\
        
        \it{MUSCAT photometry parameters}\\
        $D_{\rm MUSCAT,g}$  &Fixed   &$1$
          &MUSCAT $g$ band photometric dilution factor.\\
        $M_{\rm MUSCAT,g}$  &$\mathcal{N}$ (0\ ,\ $0.1^{2}$)   &$0.0002^{+0.0009}_{-0.0009}$
          &Mean out-of-transit flux of MUSCAT $g$ band photometry.\\
        $\sigma_{\rm MUSCAT,g}$ (ppm)  &$\mathcal{J}$ ($0.1$\ ,\ $10^{5}$)  &$27.6^{+711.3}_{-26.8}$
          &MUSCAT $g$ band additive photometric jitter term.\\
        $q_{\rm MUSCAT,g}$   &$\mathcal{N}$ ($0.79$\ ,\ $0.1^{2}$)              &$0.77^{+0.08}_{-0.08}$    &Linear limb darkening coefficient.\\
        
        $D_{\rm MUSCAT,r}$  &Fixed   &$1$
          &MUSCAT $r$ band photometric dilution factor.\\
        $M_{\rm MUSCAT,r}$  &$\mathcal{N}$ (0\ ,\ $0.1^{2}$)   &$0.0001^{+0.0003}_{-0.0003}$
          &Mean out-of-transit flux of MUSCAT $r$ band photometry.\\
        $\sigma_{\rm MUSCAT,r}$ (ppm) &$\mathcal{J}$ ($0.1$\ ,\ $10^{5}$)   &$8.96^{+177.7}_{-8.4}$
          &MUSCAT $r$ band additive photometric jitter term.\\
        $q_{\rm MUSCAT,r}$      &$\mathcal{N}$ ($0.73$\ ,\ $0.1^{2}$)          &$0.76^{+0.06}_{-0.05}$     &Linear limb darkening coefficient.\\
        
        $D_{\rm MUSCAT,z}$  &Fixed   &$1$
          &MUSCAT $z$ band photometric dilution factor.\\
        $M_{\rm MUSCAT,z}$  &$\mathcal{N}$ (0\ ,\ $0.1^{2}$)   &$0.0002^{+0.0002}_{-0.0002}$
          &Mean out-of-transit flux of MUSCAT $z$ band photometry.\\
        $\sigma_{\rm MUSCAT,z}$ (ppm) &$\mathcal{J}$ ($0.1$\ ,\ $10^{5}$)  &$11.7^{+192.4}_{-11.1}$
          &MUSCAT $z$ band additive photometric jitter term.\\
        $q_{\rm MUSCAT,z}$       &$\mathcal{N}$ ($0.44$\ ,\ $0.1^{2}$)            &$0.45^{+0.05}_{-0.05}$   &Linear limb darkening coefficient.\\
        
        \it{Stellar parameters}\\
        ${\rho}_{\ast}$ ($\rm kg\ m^{-3}$) &$\mathcal{J}$ ($100$\ ,\ $\rm 100^{2}$)   &$4278^{+412}_{-395}$
         &Stellar density.\\
        \it{RV parameters}\\
        $K_{b}$ ($\rm m\ s^{-1}$)  &$\mathcal{U}$ ($0$\ ,\ $200$)      &$66.5^{+14.1}_{-14.0}$
        &RV semi-amplitude of \tar b.\\
        $\rm \mu_{SPIRou}$ ($\rm m\ s^{-1}$) &$\mathcal{U}$ ($29300$\ ,\ $29500$)   &$29402.4^{+11.1}_{-11.5}$
        &Systemic velocity for SPIRou.\\
        $\rm \sigma_{SPIRou}$ ($\rm m\ s^{-1}$) &$\mathcal{J}$ ($0.1$\ ,\ $100$)   &$37.3^{+10.8}_{-8.4}$
        &Extra jitter term for SPIRou.\\
        \hline\hline
    \label{tranpriors}    
    \end{tabular}
\end{table*}

\begin{table}
\caption{Derived physical parameters from the final joint fit for \tar.}
 \begin{tabular}{lcr}   
    \hline\hline
        Parameter        &Best-fit           &Description\\\hline
        $R_{p}/R_{\ast}$  &$0.155^{+0.002}_{-0.002}$ &Planet radius in units of stellar radius.\\
        $R_{p}$ ($R_{J}$)  &$0.83^{+0.06}_{-0.06}$ &Planet radius.\\
        $M_{P}$ ($M_{J}$)  &$0.40^{+0.09}_{-0.10}$ &Planet mass.\\
        $\rho_{p}$ ($\rm g\ cm^{-3}$)  &$0.93^{+0.49}_{-0.35}$ &Planet density.\\
        ${b}$      &$0.33^{+0.08}_{-0.11}$ &Impact parameter.\\
        $a/R_{\ast}$     &$20.97^{+0.65}_{-0.67}$ &Semi-major axis in units of stellar radii.\\
        $a$ (AU)       &$0.052^{+0.005}_{-0.004}$ &Semi-major axis.\\
        $i$ (deg)     &$89.1^{+0.3}_{-0.3}$ &Inclination angle.\\
        $T_{\rm eq}^{[1]}$ (K)     &$565^{+28}_{-31}$ &Equilibrium temperature.\\
        \hline\hline 
    \end{tabular}
    \begin{tablenotes}
    \item[1]  [1]\ We assume there is no heat distribution between the dayside and nightside, and that the albedo is zero.
    \end{tablenotes}
    \label{physical_parameter}
\end{table}

\section{Discussion}\label{discussion}

\subsection{A lack of hot massive giant planets around M dwarfs?}
Figure \ref{aplot} shows the planet-to-star mass ratio ($q$) as a function of separation distance ($a$) of all giant planets ($0.3\ M_{J}<M_{p}<13.6\ M_{J}$) around M dwarfs detected by different methods. Regarding the microlensing sample, since most lens systems are still blended with their sources, it is hard to determine the spectral type of the host star in the lens system\footnote{Even if the lens and their sources have separated after sufficient long time due to the proper motion, it is still difficult as the host stars of the lens systems are very faint (typically $V\sim25$ mag).}. Thus we simply set a host-mass threshold between 0.08 and 0.65 $M_{\odot}$, and we filter out targets that meet the mass cut. While for the other three, we pick out the sample mainly based on the spectral information. We only consider the mass ratio here because most microlensing light-curve analyses do not provide the masses of the host and the planet, although the planet-to-host mass ratio, $q$, is well determined \citep{Mao1991,Gould1992}. To measure the mass of microlensing planet, one needs two observables \citep{Zang2020}, but most microlensing planets do not have them, and thus a Bayesian analysis is needed to estimate the host mass, which has a typical $1\sigma$ uncertainty of $\sim 0.3~M_{\odot}$. Thus, it is challenging to classify microlensing planets according to different types of host stars. However, several microlensing detections with unambiguous mass measurements demonstrate that gaint planets orbiting M dwarfs are common (e.g., \citealt{Bennett2020}).

Four giant planets identified by direct imaging that are far from their host M dwarfs are located at the high-mass-ratio region. This is likely caused by observational biases as the imaging method has difficulty to detect low mass Jupiters with $M_{p}$ around $1\ M_{J}$ (all of these four planets have $M_{p}\gtrsim10\ M_{J}$). Microlensing, however, is sensitive to all kinds of widely separated planets with masses ranging from  super-Jupiter down to Earth (e.g., \citealt{Zang2021}). A total of 55 giant planets harboured by M dwarfs have been discovered with projected separation distance $a_{\bot}\gtrsim1$ AU\footnote{The solutions of 12 microlensing systems have the so-called close-wide degeneracy, shown in pairs as translucent blue squares in Figure \ref{aplot}.}. There is a wide mass ratio distribution of those microlensing systems, most of which have $q\lesssim10^{-2}$, indicating that cold Jupiters around M dwarfs are possibly common and diverse. 

A similar trend can also been seen in the RV-only sample whose separation distances are between 0.1 and 10 AU, although RV can only determine the minimum mass ratio $q_{\rm min}$ for those non-transiting systems. Currently, there are no RV-only giant planets with $q_{\rm min}\geq10^{-2}$ that have been detected around M dwarfs, which is likely due to observational biases as follows. Unlike microlensing, which is not limited by the lens flux, determining the companion mass spectroscopically requires central stars to be relatively bright (typically $V<13$ mag). Thus the RV-only sample may miss giant planets around faint late-type M dwarfs, which have higher $q_{\rm min}$ compared with equivalent planets around early-type M dwarfs. For massive early-type M dwarfs, however, some of their companions within that mass ratio range should belong to brown dwarfs, which are not included here. Furthermore, no giant planets have been detected within 0.1 AU of their host M dwarfs from RV-only surveys. This phenomenon can be attributed to the RV observational strategy. Most RV surveys focus on bright nearby M dwarfs and the total sample size is small (roughly $\sim200$). Thus it is reasonable to find none RV-only giant planets in this region given the low occurrence rate of hot Jupiter ($\sim0.5\%$, \citealt{Fressin2013,Petigura2018,Zhou2019}). Therefore, deep transit surveys play a crucial role in detecting such candidates as they are sensitive to planets with small semi-major axis and large mass ratio, which also more likely to have a large planet-to-star radius ratio. 

Interestingly, all five known transiting giant planet systems and \tar b turn out to have small mass ratio $q\sim10^{-3}$ and there is a possible dearth at the region with $q>2\times10^{-3}$ and $a<0.1$ AU (see the shaded region in Figure \ref{aplot}). Part of that may result from the flux-limit problem above (for $q\geq10^{-2}$). We note that this deficiency feature may reflect a more fundamental link to the planet formation theory. Recent work from \cite{Liu2019} constructed a pebble-driven planet population synthesis model, and their simulation results suggest that gas giants may mainly form when the central stars are more massive than $0.3\ M_{\odot}$ (see Figure 7 in \citealt{Liu2019}). This is because planets stop increasing their core masses when they reach the pebble isolation mass $M_{\rm iso}$, which is proportional to the stellar mass as $M_{\rm iso}\propto M_{\ast}^{4/3}$. Following gas accretion onto planets with small $M_{\rm iso} $ is limited due to a slow Kelvin–Helmholtz contraction. Thus they would stop before the runaway gas accretion and be left as rock- or ice-dominated planets with tiny atmospheres. If this is the case, we then expect few giant planets with relatively high mass ratio above $10^{-3}$ when their host masses are below $0.3\ M_{\odot}$. Note that the known ``brown dwarf desert'' ($35 \leq M\sin i \leq 55\ M_{J}$ and orbital periods under 100 days), studied by \cite{Ma2014} using all available data of close brown dwarfs around solar-type stars, is also located in this region with $q\gtrsim0.05$. This lower limit is estimated based on the lower limit of the brown dwarf desert $35\ M_{J}$ and the typical mass upper limit of M dwarfs $0.65\ M_{\odot}$. However, it is still unclear whether the deficiency between $2\times10^{-3}$ and $10^{-2}$ in mass ratio is physical. Compared with the known planets (red squares plus \tar b in Figure \ref{aplot}), the transit method is, in principle, more sensitive to giant planets with a larger mass ratio (i.e., larger radius ratio) located in this deficiency region. Additionally, if detected by transit survey, planet candidates within this parameter space range should be easily confirmed by the RV method. Due to the lack of transiting giant planets around M dwarfs, we cannot draw any conclusions yet. Hopefully, the \tess\ QLP Faint Star ($10.5<T<13.5$ mag) Search could provide more such systems and check if this depletion feature is real (Kunimoto et al, in prep).

\begin{figure}
\centering
\includegraphics[width=0.49\textwidth]{a_q.pdf}
\caption{Planet-to-star mass ratio of all giant planets around M dwarfs as a function of semi-major axis. Different colors represent planets detected by different methods. The two horizontal black dashed lines represent the selection threshold of our sample. The upper limit is $q=13.6\ M_{J}/0.08\ M_{\odot}=0.17$, while the lower limit corresponds to $q=0.3\ M_{J}/0.65\ M_{\odot}=4.6\times10^{-4}$. \tar b is marked as a magenta circle. The grey shaded region represents a possible paucity of hot massive giant planets around M dwarfs.} 
\label{aplot}
\end{figure}

\subsection{Metallicity Dependence}\label{metallicity_dependence}
Although the core accretion theory \citep{Pollack1996} has predicted rare giant planets around M dwarfs due to their low protoplanetary disk mass as $M_{\rm disk}$ linearly scales with the stellar mass $M_{\ast}$ \citep{Andrews2013}, this defect may be compensated if the parent star is metal rich, which could theoretically supply more solid material used for accretion. Alternatively, gravitational instability (GI) could also form giant planets around M dwarfs \citep{Boss2006}, though simulation work from \cite{Cai2006} suggested that GI is unlikely the major mechanism that produce most observed planets. Under the GI hypothesis, we expect that there would not exist a strong dependence between giant planet formation and host metallicity and that even stars with relatively low metallicity should harbour gas giants \citep{Boss2002}.

\begin{figure}
\centering
\includegraphics[width=0.49\textwidth]{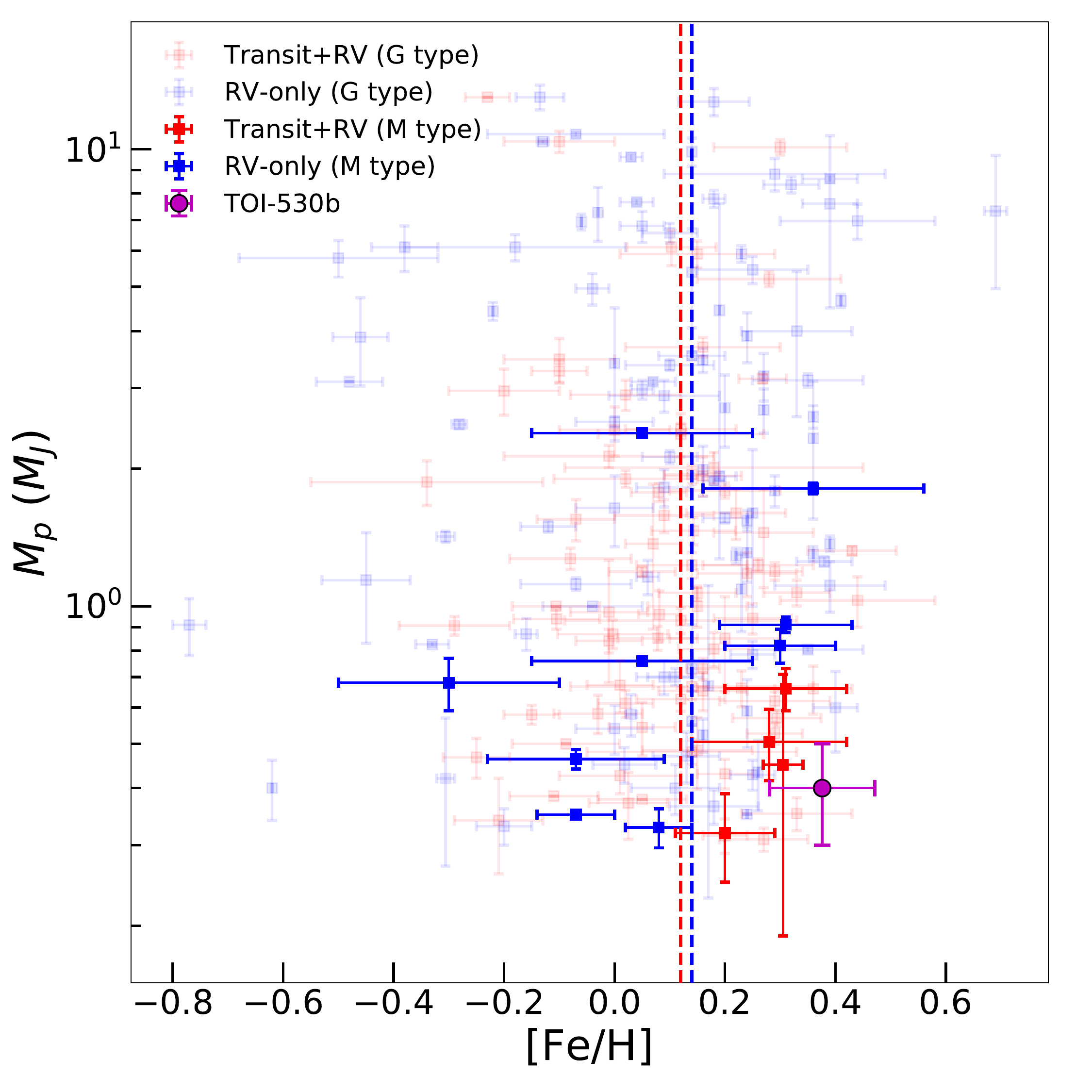}
\caption{Planet mass $M_{p}$ vs. host star metallicity. Different colors represent planets detected by different methods (transit/RV). The red and blue dashed lines are the weighted-mean metallicity of solar-type stars hosting giant planets detected by transit and RV-only observations. \tar b is marked as a magenta circle. All hot giant planet M-dwarf hosts tend to have a higher metallicity than G dwarf hosts (see Section \ref{metallicity_dependence}).} 
\label{fehplot}
\end{figure}

In order to investigate the metallicity dependence observationally, we retrieve a list of solar-type stars (simply selected based on $0.90\ M_{\odot}<M_{\ast}<1.06\ M_{\odot}$) hosting giant planets from the NASA Exoplanet Archive \citep{Akeson2013}. We find a total of 88 transiting and 102 RV-only systems. Most transiting giant planets are hot with semi-major axis $a\lesssim 0.1$ AU, while the majority of RV-only planets are cold with semi-major axes beyond  $0.1$ AU. Figure~\ref{fehplot} illustrates the metallicity distribution of their hosts, indicating that hot and cold giant planets do not present much difference in [Fe/H] preference around solar-type stars (see the transparent points in Figure \ref{fehplot}). The weighted-mean metallicity of both transiting and RV-only giant planet central stars is above the solar value but almost the same: 0.12 and 0.14, respectively. This is consistent with the conclusions from early work suggesting that the frequency of giant planets increases with stellar metallicity \citep{Santos2004,Fischer2005,Sousa2011}. However, it is not the same for gas giants around M dwarfs. 

Among all giant planets around M dwarfs, we only find 4 transiting and 9 RV-only systems that have metallicity measurements in the literature. For cold (RV-only) giant planets, the metallicities of their M-dwarf hosts are distributed on both sides of the median value 0.14 (the green dashed line in Figure \ref{fehplot}) seen for the solar-type stars, though the uncertainties are large. Some of them are formed around metal poor M stars (e.g., GJ 832b, \citealt{Bailey2009}). This is plausibly in agreement with the previous finding that clump formation is fairly insensitive to the metallicity of the parent star \citep{Boss2002}, implying that part of the formation of cold giant planets around M dwarfs may take place through GI. Indeed, the recently detected planetary system GJ 3512b, whose host star has a solar metallicity $-0.07\pm0.16$, favours the GI formation scenario \citep{Morales2019}.

However, the host stars of four known transiting giant planets together with \tar\ tend to be metal rich, all of which have [Fe/H] higher than the aforementioned median value 0.12 (the red dashed line in Figure \ref{fehplot}). It indicates that the formation of hot (transiting) giant planets around M dwarfs may have a strong dependence on the host metallicity as predicted by the core accretion theory. This is consistent with the recent findings from \cite{Maldonado2020} (see the left panel of their Figure 14), which reveals a correlation between the metallicities of M dwarfs and their probability of hosting giant planets. Compared with hot giant planets around solar-type stars, the formation of hot giant planets around M dwarfs possibly requires higher host metallicity, though the number of detections is probably too small to confirm this claim. To examine the statistical significance of this feature, we carry out a Kolmogorov-Smirnov (K-S) test. We calculate the K-S statistic between the metallicities of stars in the G- and M-type transiting sample, which yields a $p$ value of 0.023. It roughly corresponds to the $2.5\sigma$ significance level, at which we can reject the null hypothesis that two samples are from the same distribution. We then adopt the bootstrap method to randomly draw distributions from the G-type parent sample and compute the K-S statistic between these random distributions and the M-type transiting sample. We repeat this procedure for 10,000 times and derive the corresponding $p$ value distribution. We find that 88\% of the total random samples have $p$ values less than 0.05 ($2\sigma$ level) while only 0.3\% of them have $p$ values less than 0.003 ($3\sigma$ level), indicating a marginal correlation. Future detections of more hot giant planets around M dwarfs will reveal whether this metallicity preference is significant or not. 





\subsection{Prospects for future observations}
Due to the faintness of the host star, high precision radial velocity spectrographs on large telescopes like the red-optical MAROON-X \citep{Seifahrt2018} or near-infrared instruments like InfrRed Doppler spectrograph (IRD; \citealt{Kotani2018}) are required to reach sufficient signal-to-noise ratio. The host star is quiet without strong intrinsic stellar activity as there are no significant flux variations in the \tess\ light curve, making it a suitable target for precise RV follow-up observations. Though a potential 9.4 d modulation signal is shown up in the \tess\ light curve, this is probably not linked to the stellar rotation given our ZTF results and previous findings from \cite{Newton2018} (see Section \ref{stellar_properties}). 

Measuring the stellar obliquity of M dwarfs hosting hot giant planets could provide clues about their origins and gain insights into their migration history. To probe the potential opportunities to observe the Rossiter-McLaughlin effect \citep{Rossiter1924,McLaughlin1924} of \tar\ and measure the projected
angle between the planet orbital and stellar equatorial planes, we estimate the RM semi-amplitude of this system as:
\begin{equation}
    A_{\rm RM} \simeq \frac{2}{3}(R_{p}/R_{\ast})^{2}\sqrt{1-b^{2}}\times v \sin i,
\end{equation}
where $b$ is the impact parameter and $v \sin i$ is the projected stellar equatorial rotation velocity. Taking the best-fit values from the light curve modeling, assuming a rotational period of 40 d for a $0.5\ M_{\odot}$ star (see Figure 4 in \citealt{Newton2018}) with an upper and lower limit of 100 and 30 d, we find $A_{\rm RM}\sim 10^{+4}_{-6}$ m/s, making the RM signal detectable with near-infrared RV observations.

Finally, we investigate the atmospheric characterization possibility by calculating the Transmission Spectroscopy Metric (TSM; \citealt{Kempton2018}) of \tar b. We obtain a TSM of $41^{+20}_{-13}$, which is much smaller than the recommended threshold of 90 for high-quality sub-Jovians ($4.0<R_{p}<10\ R_{\oplus}$) targets from \cite{Kempton2018}. Thus we conclude that \tar b is not a promising target for future atmospheric composition studies. 


\section{Summary and Conclusions}\label{conclusion}
In this paper, we present the discovery and characterization of a transiting giant planet \tar b around an early, metal-rich M dwarf. Space and ground photometry as well as SPIRou RVs reveal that \tar b has a radius of $0.83\pm0.05\ R_{J}$ and a mass of $0.4\pm0.1\ M_{J}$ on a 6.39-d orbit. Although it is challenging to probe the atmospheric properties of \tar b due to the faintness of the host star, \tar\ is still a suitable target to study the stellar obliquity. Furthermore, we report a potential paucity of hot massive giant planets around M dwarfs with separation distance smaller than $0.1$ AU and planet-to-star mass ratio between $2\times10^{-3}$ and $10^{-2}$. We also identify a possible correlation between hot giant planet formation and the metallicity of its parent M dwarf. However, due to the current small sample of such systems, we could not make any firm conclusions in this context. Future near-infrared spectroscopic surveys such as SPIRou Legacy Survey-Planet Search (SLS-PS; \citealt{Moutou2017,Fouque2018}) shall remedy this situation.

\section*{Affiliations}
$^{1}$Department of Astronomy, Tsinghua University, Beijing 100084, People's Republic of China\\
$^{2}$Department of Physics, Tsinghua University, Beijing 100084, People's Republic of China\\
$^{3}$National Astronomical Observatories, Chinese Academy of Sciences, 20A Datun Road, Chaoyang District, Beijing 100012, People's Republic of China\\
$^{4}$Canada-France-Hawaii Telescope, CNRS, Kamuela, HI 96743, USA\\
$^{5}$Univ. de Toulouse, CNRS, IRAP, 14 Avenue Belin, 31400 Toulouse, France\\
$^{6}$Department of Physics and Astronomy, Vanderbilt University, 6301 Stevenson Center Ln., Nashville, TN 37235, USA\\
$^{7}$Department of Physics, Fisk University, 1000 17th Avenue North, Nashville, TN 37208, USA\\
$^{8}$Department of Astronomy, University of California Berkeley, Berkeley, CA 94720, USA\\
$^{9}$Komaba Institute for Science, The University of Tokyo, 3-8-1 Komaba, Meguro, Tokyo 153-8902, Japan\\
$^{10}$Instituto de Astrof\'isica de Canarias (IAC), V\'ia L\'actea s/n, E-38205 La Laguna, Tenerife, Spain\\
$^{11}$Dept. Astrof\'sica, Universidad de La Laguna (ULL), E-38206 La Laguna, Tenerife, Spain\\
$^{12}$NASA Exoplanet Science Institute, Caltech/IPAC, Mail Code 100-22, 1200 E. California Blvd., Pasadena, CA 91125, USA\\
$^{13}$NASA Ames Research Center, Moffett Field, CA 94035, USA\\
$^{14}$Center for Astrophysics ${\rm \mid}$ Harvard {\rm \&} Smithsonian, 60 Garden Street, Cambridge, MA 02138, USA\\
$^{15}$Department of Physics and Kavli Institute for Astrophysics and Space Research, Massachusetts Institute of Technology, Cambridge, MA 02139, USA\\
$^{16}$NASA Goddard Space Flight Center, 8800 Greenbelt Road, Greenbelt, MD 20771, USA\\
$^{17}$University of Maryland, Baltimore County, 1000 Hilltop Circle, Baltimore, MD 21250, USA\\
$^{18}$Department of Physics \& Astronomy, University of Kansas, 1082 Malott,1251 Wescoe Hall Dr., Lawrence, KS 66045, USA\\
$^{19}$Embry-Riddle Aeronautical University, Prescott, AZ\\
$^{20}$El Sauce Observatory, Coquimbo Province, Chile\\
$^{21}$Department of Astronomy and Astrophysics, University of California, Santa Cruz, CA 95064, USA\\
$^{22}$Department of Astronomy, California Institute of Technology, Pasadena, CA 91125, USA\\
$^{23}$Okayama Observatory, Kyoto University, 3037-5 Honjo, Kamogatacho, Asakuchi, Okayama 719-0232, Japan\\
$^{24}$Department of Multi-Disciplinary Sciences, Graduate School of Arts and Sciences, The University of Tokyo, 3-8-1 Komaba, Meguro, Tokyo 153-8902, Japan\\
$^{25}$Department of Earth and Planetary Science, Graduate School of Science, The University of Tokyo, 7-3-1 Hongo, Bunkyo-ku, Tokyo 113-0033, Japan\\
$^{26}$Zhejiang Institute of Modern Physics, Department of Physics \& Zhejiang University-Purple Mountain Observatory Joint Research Center for Astronomy, Zhejiang University, Hangzhou 310027, China\\
$^{27}$Department of Astronomy, The University of Tokyo, 7-3-1 Hongo, Bunkyo-ku, Tokyo 113-0033, Japan\\
$^{28}$U.S. Naval Observatory, Washington, D.C. 20392, USA\\
$^{29}$Japan Science and Technology Agency, PRESTO, 3-8-1 Komaba, Meguro, Tokyo 153-8902, Japan\\
$^{30}$Astrobiology Center, 2-21-1 Osawa, Mitaka, Tokyo 181-8588, Japan\\
$^{31}$Department of Earth, Atmospheric and Planetary Science, Massachusetts Institute of Technology, 77 Massachusetts Avenue, Cambridge, MA 02139, USA\\
$^{32}$Space Telescope Science Institute, 3700 San Martin Drive, Baltimore, MD, 21218, USA\\
$^{33}$Department of Aeronautics and Astronautics, MIT, 77 Massachusetts Avenue, Cambridge, MA 02139, USA\\
$^{34}$Department of Astrophysical Sciences, Princeton University, 4 Ivy Lane, Princeton, NJ 08544, USA\\
$^{\ast}$51 Pegasi b Fellow\\
\section*{Acknowledgements}
We are grateful to Coel Hellier for the insights regarding the WASP data. We thank Elisabeth Newton, Robert Wells, Hongjing Yang and Weicheng Zang for useful discussions. We also thank Elise Furlan for the contributions to the speckle data and Nadine Manset for scheduling the SPIRou observations.
This work is partly supported by the National Science Foundation of China (Grant No. 11390372 and 11761131004 to SM and TG). 
This research uses data obtained through the Telescope Access Program (TAP), which has been funded by the TAP member institutes.
This work is partly supported by JSPS KAKENHI Grant Numbers JP17H04574 , JP18H05439, 20K14521, JST PRESTO Grant Number JPMJPR1775, and the Astrobiology Center of National Institutes of Natural Sciences (NINS) (Grant Number AB031010).
This article is based on observations made with the MuSCAT2 instrument, developed by ABC, at Telescopio Carlos Sánchez operated on the island of Tenerife by the IAC in the Spanish Observatorio del Teide.
Some of the observations in the paper made use of the High-Resolution Imaging instrument ‘Alopeke obtained under LLP GN-2021A-LP-105. ‘Alopeke was funded by the NASA Exoplanet Exploration Program and built at the NASA Ames Research Center by Steve B. Howell, Nic Scott, Elliott P. Horch, and Emmett Quigley. Data were reduced using a software pipeline originally written by Elliott Horch and Mark Everett. ‘Alopeke was mounted on the Gemini North telescope of the international Gemini Observatory, a program of NSF’s OIR Lab, which is managed by the Association of Universities for Research in Astronomy (AURA) under a cooperative agreement with the National Science Foundation. on behalf of the Gemini partnership: the National Science Foundation (United States), National Research Council (Canada), Agencia Nacional de Investigación y Desarrollo (Chile), Ministerio de Ciencia, Tecnología e Innovación (Argentina), Ministério da Ciência, Tecnologia, Inovações e Comunicações (Brazil), and Korea Astronomy and Space Science Institute (Republic of Korea). 
Funding for the TESS mission is provided by NASA's Science Mission directorate. 
We acknowledge the use of \tess\ public data from pipelines at the \tess\ Science Office and at the \tess\ Science Processing Operations Center. 
Resources supporting this work were provided by the NASA High-End Computing (HEC) Program through the NASA Advanced Supercomputing (NAS) Division at Ames Research Center for the production of the SPOC data products.
This research has made use of the Exoplanet Follow-up Observation Program website, which is operated by the California Institute of Technology, under contract with the National Aeronautics and Space Administration under the Exoplanet Exploration Program. 
This paper includes data collected by the \tess\ mission, which are publicly available from the Mikulski Archive for Space Telescopes\ (MAST). 
This work has made use of data from the European Space Agency (ESA) mission
{\it Gaia} (\url{https://www.cosmos.esa.int/gaia}), processed by the {\it Gaia} Data Processing and Analysis Consortium (DPAC,
\url{https://www.cosmos.esa.int/web/gaia/dpac/consortium}). Funding for the DPAC has been provided by national institutions, in particular the institutions participating in the {\it Gaia} Multilateral Agreement.
This work made use of \texttt{tpfplotter} by J. Lillo-Box (publicly available in www.github.com/jlillo/tpfplotter), which also made use of the python packages \texttt{astropy}, \texttt{lightkurve}, \texttt{matplotlib} and \texttt{numpy}.
\section*{Data Availability}
This paper includes photometric data collected by the \tess\ mission and ground instruments, which are publicly available in ExoFOP, at \url{https://exofop.ipac.caltech.edu/tess/target.php?id=387690507}. All spectroscopy data underlying this article are listed in the text. All of the high-resolution speckle imaging data is available at the NASA exoplanet Archive with no proprietary period.



\bibliographystyle{mnras}
\bibliography{planet} 




\appendix
\section{Prior settings for TESS-only fit, ground photometric data detrending and RV-only modeling}

\begin{table*}
    \centering
    \caption{Prior settings and posterior values for the fit to the \tess\ only data.}
    \begin{tabular}{lccr}
        \hline\hline
        Parameter       &Best-fit Value       &Prior     &Description\\\hline
        \it{Planetary parameters}\\
        $P_{b}$ (days)   &$6.38758^{+0.00003}_{-0.00003}$  
        &$\mathcal{U}$ ($6.2$\ ,\ $6.6$)
        &Orbital period of \tar b.\\
        $T_{0,b}$ (BJD-2457000)    &$1470.201^{+0.002}_{-0.003}$ 
        &$\mathcal{U}$ ($1468$\ ,\ $1472$) 
        &Mid-transit time of \tar b.\\
        $r_{1,b}$    &$0.436^{+0.078}_{-0.070}$ 
        &$\mathcal{U}$ (0\ ,\ 1)
        &Parametrisation for {\it p} and {\it b}.\\
        $r_{2,b}$    &$0.152^{+0.004}_{-0.004}$ 
        &$\mathcal{U}$ (0\ ,\ 1)
        &Parametrisation for {\it p} and {\it b}.\\
        $e_{b}$                     &0  &Fixed  &Orbital eccentricity of \tar b.\\
        $\omega_{b}$ (deg)          &90 &Fixed  &Argument of periapsis of \tar b.\\
        \it{Stellar parameters}\\
        ${\rho}_{\ast}$ ($\rm kg\ m^{-3}$)   &$4702^{+207}_{-206}$
        &$\mathcal{J}$ ($100$\ ,\ $\rm 100^{2}$) &Stellar density.\\
        \it{\tess\ photometry parameters}\\
        $D_{\rm TESS}$     &$1$ 
        &Fixed      &\tess\ photometric dilution factor.\\
        $M_{\rm TESS}$    &$-0.000009^{+0.00009}_{-0.00009}$
        &$\mathcal{N}$ (0\ ,\ $0.1^{2}$)      &Mean out-of-transit flux of \tess\ photometry.\\
        $\sigma_{\rm TESS}$ (ppm) &$0.03^{+15.11}_{-0.02}$
        &$\mathcal{J}$ ($10^{-6}$\ ,\ $10^{6}$)      &\tess\ additive photometric jitter term.\\
        $q_{1}$                &$0.31^{+0.28}_{-0.17}$       &$\mathcal{U}$ (0\ ,\ 1)  &Quadratic limb darkening coefficient.\\
        $q_{2}$                &$0.42^{+0.32}_{-0.25}$       &$\mathcal{U}$ (0\ ,\ 1)  &Quadratic limb darkening coefficient.\\
        
        \hline\hline 
    \end{tabular}
    \label{tess_only_fit_priors}
\end{table*}

\begin{table*}
    \centering
    \caption{Prior settings for detrending the ground data.}
    \begin{tabular}{lcr}
        \hline\hline
        Parameter          &Prior     &Description\\\hline
        \it{Planetary parameters}\\
        $P_{b}$ (days)     
        &$\mathcal{U}$ (6.386\ ,\ 6.388)
        &Orbital period of \tar b.\\
        $T_{0,b}$ (BJD-2457000)   
        &$\mathcal{U}$ ($1470.198$\ ,\ $1470.204$) 
        &Mid-transit time of \tar b.\\
        $r_{1,b}$     
        &$\mathcal{U}$ (0.4\ ,\ 0.7)
        &Parametrisation for {\it p} and {\it b}.\\
        $r_{2,b}$    
        &$\mathcal{U}$ (0.13\ ,\ 0.17)
        &Parametrisation for {\it p} and {\it b}.\\
        $e_{b}$                       &0 (Fixed)  &Orbital eccentricity of \tar b.\\
        $\omega_{b}$ (deg)            &90 (Fixed)  &Argument of periapsis of \tar b.\\
        \it{Stellar parameters}\\
        ${\rho}_{\ast}$ ($\rm kg\ m^{-3}$)   
        &$\mathcal{N}$ ($4702$\ ,\ $\rm 207^{2}$) &Stellar density.\\
        \it{Photometry parameters for each ground light curve}\\
        $D_{i}$      
        &1 (Fixed)      &Photometric dilution factor.\\
        $M_{i}$    
        &$\mathcal{N}$ (0\ ,\ $0.1^{2}$)      &Mean out-of-transit flux of ground photometry.\\
        $\sigma_{i}$ (ppm) 
        &$\mathcal{J}$ ($10^{-1}$\ ,\ $10^{5}$)      &Ground additive photometric jitter term.\\
        $q_{i}$                       &$\mathcal{U}$ (0\ ,\ 1)  &Linear limb darkening coefficient.\\
        
        \hline\hline 
    \end{tabular}
    \label{ground_only_fit_priors}
\end{table*}

\begin{table*}
    \centering
    \caption{Prior settings and posteriors of RV-only modeling}
    \begin{tabular}{lccr}
        \hline\hline
        Parameter       &Priors         &Best-fit &Description   \\\hline
        \it{Planetary parameters}\\
        $P_{b}$ (days)   &$6.38758$ (Fixed) &$6.38758$  
         &Orbital period of \tar b.\\
        
        $T_{0,b}$ (BJD)   &$1470.201$ (Fixed) 
        &$1470.201$ 
        &Mid-transit time of \tar b.\\
        
        $e$ &$0$ (Fixed)  &$0$  &Orbital eccentricity of \tar b. \\
        $\omega$ &$90$ (Fixed)  &$90$ &Argument of periapsis of \tar b. \\
        
        \it{RV parameters}\\
        $\rm \mu_{SPIRou}$ ($\rm m\ s^{-1}$) &$\mathcal{U}$ ($29300$\ ,\ $29500$) &$29403.0^{+11.7}_{-12.4}$ &Systemic velocity for SPIRou.\\ 
        $\rm \sigma_{SPIRou}$ ($\rm m\ s^{-1}$) &$\mathcal{J}$ ($0.1$\ ,\ $100$) &$37.5^{+11.2}_{-8.9}$ &Extra jitter term for SPIRou.\\

        $K_{b}$ ($\rm m\ s^{-1}$)   &$\mathcal{U}$ ($0$\ ,\ $200$)
        &$67.2^{+15.1}_{-14.4}$ &RV semi-amplitude of TOI-530b.\\
        \hline
        \hline\hline 
    \end{tabular}
    \label{rvonly_priors}
\end{table*}

\bsp	
\label{lastpage}
\end{document}